\documentclass[twocolumn,english,aps,arxiv,superscriptaddress]{revtex4-1}
\usepackage[utf8]{inputenc}
\usepackage{natbib}
\usepackage{grffile}
\usepackage{gensymb}
\usepackage{color}
\usepackage{babel}
\usepackage{amsmath}
\usepackage{amsfonts}
\usepackage{graphicx}
\usepackage{epstopdf}
\usepackage{multirow}
\usepackage{cancel}
\usepackage{lineno}
\PassOptionsToPackage{normalem}{ulem}
\usepackage{ulem}
\usepackage[unicode=true, bookmarks=false, breaklinks=false,pdfborder={0 0 1},colorlinks=false] {hyperref}
\usepackage{breakurl}

\begin{document}

\title{Determining the Nature of Magnetism in Altermagnetic Candidate RuO$_2$}

\author{Tiema Qian}
\affiliation{National High Magnetic Field Laboratory, Los Alamos National Laboratory, Los Alamos, New Mexico 87545, USA}

\author{Aya Rutherford}
\affiliation{Department of Physics and Astronomy,University of Tennessee, Knoxville, Tennessee 37996, USA}

\author{Eun Sang Choi}
\affiliation{National High Magnetic Field Laboratory, Tallahassee, Florida 32310, USA}

\author{Haidong Zhou}
\affiliation{Department of Physics and Astronomy,University of Tennessee, Knoxville, Tennessee 37996, USA}

\author{Boris Maiorov}
\affiliation{National High Magnetic Field Laboratory, Los Alamos National Laboratory, Los Alamos, New Mexico 87545, USA}

\author{Minseong Lee}
\affiliation{National High Magnetic Field Laboratory, Los Alamos National Laboratory, Los Alamos, New Mexico 87545, USA}

\author{Christopher A. Mizzi}
\email{Corresponding author: mizzi@lanl.gov}
\affiliation{National High Magnetic Field Laboratory, Los Alamos National Laboratory, Los Alamos, New Mexico 87545, USA}

\begin{abstract}
\noindent The terminology ``altermagnetism'' has recently been adopted to describe collinear magnetic order with no net magnetization and non-relativisitic, momentum-dependent spin-splitting. The archetypal material used to theoretically explore altermagnetism is RuO$_2$, but there has been significant debate as to whether RuO$_2$ possesses magnetic, let alone altermagnetic, order. To address questions surrounding the nature of magnetism in RuO$_2$, we combine symmetry-sensitive torque magnetometry and magnetization measurements in single crystals. The data are inconsistent with collinear magnetic order possessing a N\'eel vector along the $c-$axis. Torque magnetometry further demonstrates an isotropic magnetic susceptibility within the $ab-$plane, indicative of neither a N\'eel vector within the $ab-$plane nor a field-induced N\'eel vector reorientation. Magnetic quantum oscillations from both techniques reveal a nearly spherical Fermi surface pocket at the Brillouin zone center, in agreement with paramagnetic electronic structure calculations. Taken together, these data indicate no detectable long-range magnetic order and, by extension, suggest no altermagnetism in high-quality RuO$_2$ single crystals.
\end{abstract}

\date{\today}
\maketitle

Collinear magnetic systems have historically been classified as ferromagnets or antiferromagnets according to the magnetic moment density averaged over a magnetic unit cell~\cite{BorovikRomanov2013}. Ferromagnets, with their finite magnetization and uniform spin-splitting, have seen widespread application owing to their easy manipulation, but have size and speed limitations~\cite{Hirohata2020}. In contrast, antiferromagnets characterized by compensated magnetization and spin-degenerate electronic states have fast ($\sim$THz) responses but are more difficult to manipulate~\cite{Jungwirth2016}. Recent analyses~\cite{vsmejkal2022beyond, vsmejkal2022emerging} based upon spin group theory~\cite{Litvin1974} have led to the identification of a third class of collinear magnetic order that has zero net magnetization (like antiferromagnets) and spin-splitting (like ferromagnets). These materials, often called ``altermagnets"~\cite{vsmejkal2022beyond, vsmejkal2022emerging}, are characterized by multipolar order parameters~\cite{Bhowal2024, McClarty2024} and momentum-dependent spin-splitting ~\cite{Hayami2019, Yuan2020} arising from the spatial symmetry operations relating opposite spin-sublattices. Importantly, spin-splitting in altermagnets is often large ($\sim0.01-1$ eV) owing to its non-relativistic origin, making this class of materials promising for magnetic sensors and scalable data storage~\cite{vsmejkal2022beyond, vsmejkal2022emerging}. Furthermore, altermagnets may serve as platforms to realize novel phases such as those associated with Fermi liquid instabilities~\cite{jungwirth2024, jungwirth2025} and $p$-wave magnets~\cite{hellenes2024pwavemagnets}.

A large body of research has emerged over the last few years examining the theoretical implications of systems exhibiting non-relativistic, momentum-dependent spin-splitting \cite{vsmejkal2022beyond, vsmejkal2022emerging, cheong2024altermagnetism, yuan2024nonrelativistic} and the experimental verification of altermagnetic properties \cite{Xu_2025, reimers2024direct, aoyama2024piezomagnetic, lee2024broken, feng2022anomalous, leiviska2024anisotropy}. Among the most studied materials in these contexts is RuO$_2$. Though it was first synthesized in the 1960s \cite{schafer1963chemie} and later developed as an electrode coating, catalyst, and low temperature thermometer \cite{kameyama1994surface, bat1995design, over2000atomic}, RuO$_2$ has seen a resurgence of interest owing to recent neutron diffraction measurements \cite{berlijn2017itinerant} and its status as the archetypal altermagnet~\cite{vsmejkal2022beyond,vsmejkal2022emerging}. The latter stems from its relatively simple crystal structure (rutile with crystallographic space group $P4_2/mnm$) and predictions of large spin-splitting~\cite{vsmejkal2022beyond,vsmejkal2022emerging}. 

Despite serving as the prevalent theoretical model for illustrating altermagnetism and exploring altermagnetic properties, a crucial question remains: is RuO$_2$ magnetically ordered? Measurements on single crystals including heat capacity, electrical transport, magnetization, and thermal expansion do not observe phase transitions associated with the onset of magnetic order \cite{cordfunke1989thermophysical, o1997gibbs, ryden1970magnetic, fletcher1968magnetic, lin2004low, PhysRevB.49.7107, touloukian1977thermophysical}. Furthermore, muon spin rotation, angle-resolved photoemission, and infrared spectroscopies point to a paramagnetic ground state~\cite{hiraishi2024nonmagnetic, kessler2024absence, pawula2024multiband, liu2024absence, wenzel2025fermi}. However, both neutron and X-ray experiments on bulk crystals have reported signatures of long-range antiferromagnetic order with small ($\sim$0.05$\mu_B$) magnetic moments \cite{zhu2019anomalous, berlijn2017itinerant}. There have been similar amounts of disagreement on magnetism in RuO$_2$ thin films \cite{feng2022anomalous, tschirner2023saturation, jeong2025metallicity, karube2022observation,bai2023efficient,  feng2024incommensurate, weber2024all, jeong2025altermagnetic, zhu2019anomalous}. 

Symmetry offers a unified framework to assess evidence for and against magnetic order in RuO$_2$ because different forms of magnetic order impose distinct, measurable constraints on materials properties. In this Letter, we combine such symmetry arguments with thermodynamic measurements in high magnetic fields to constrain the allowed forms of magnetic order in high-quality RuO$_2$ single crystals. Leveraging the sensitivity of torque magnetometry, a thermodynamic probe of magnetic susceptibility anisotropy, we determine that only magnetic order characterized by an isotropic magnetic susceptibility within the $ab-$plane and a larger susceptibility along the $c-$axis are consistent with experiment. To distinguish between the remaining possible forms of magnetic order, we combine these symmetry constraints with quantitative arguments derived from magnetization measurements which indicate itinerant paramagnetism. We confirm this interpretation by demonstrating close correspondence between the theoretical Fermi surface of paramagnetic RuO$_2$ and the experimental Fermi surface determined from magnetic quantum oscillations.

\begin{figure}
    \centering
    \includegraphics[width=3.2in, clip]{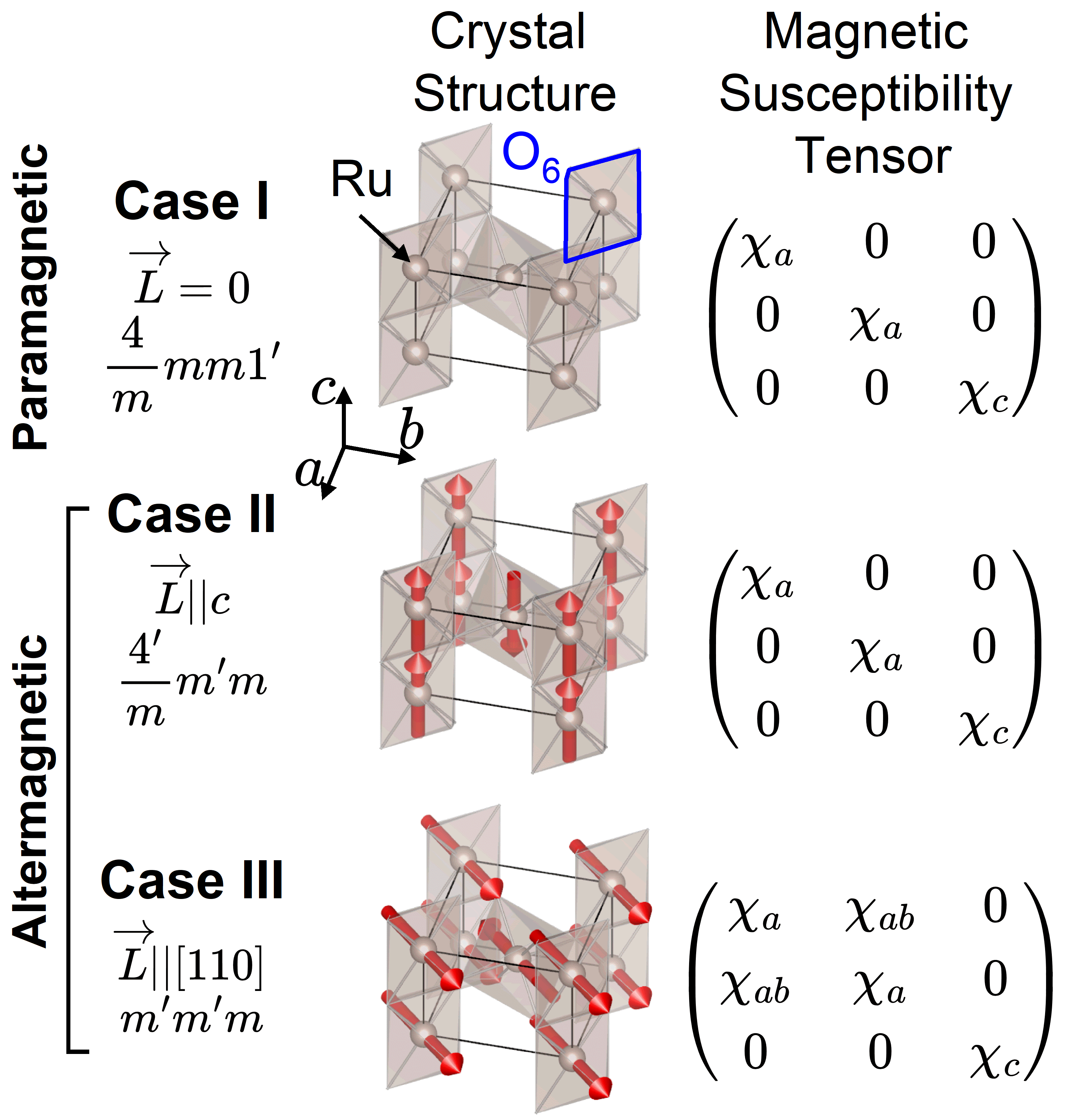}
    \caption{Three possible magnetic states in RuO$_2$. Case I is paramagnetic and Cases II-III are different altermagnetic orders. The direction of the N\'eel vector ($\vec{L}$), magnetic point group, crystal structure, and magnetic susceptibility tensor are provided for each case. Tensors are given with respect to the tetragonal axes of the parent paramagnetic structure.}
    \label{intro}
\end{figure}{}

In Fig.~\ref{intro} we examine the form of the magnetic susceptibility tensor (which describes the magnetization arising from a magnetic field to lowest order) for the rutile crystal structure with different magnetic orders~\cite{PerezMato2015}. Scenarios in which a rutile crystal is either paramagnetic (Case I) or altermagnetic with the N\'eel vector along the $c-$axis (Case II) allow for uniaxial magnetic anisotropy with an isotropic magnetic response within the $ab-$plane ($\chi_a$) and a distinct response along the $c-$axis ($\chi_c$). Case II is the most commonly reported magnetically ordered ground state for RuO$_2$ (\textit{e.g.},~\cite{berlijn2017itinerant,feng2022anomalous}). Magnetic susceptibility has the same form in Cases I and II because a N\'eel vector along the $c-$axis maintains a tetragonal system. If the N\'eel vector was in a direction other than the $c-$axis, the symmetry of the system would be lower than tetragonal allowing for additional independent magnetic susceptibility components. One such example is Case III, where the N\'eel vector is along [110]. Case III has been reported for RuO$_2$ thin films \cite{feng2022anomalous, tschirner2023saturation} and CoF$_2$ single crystals \cite{Bazan1975weak} at high magnetic fields. Other instances are considered in SM \cite{SM}.

\begin{figure*}
    \centering
    \includegraphics[width=6.5in, clip]{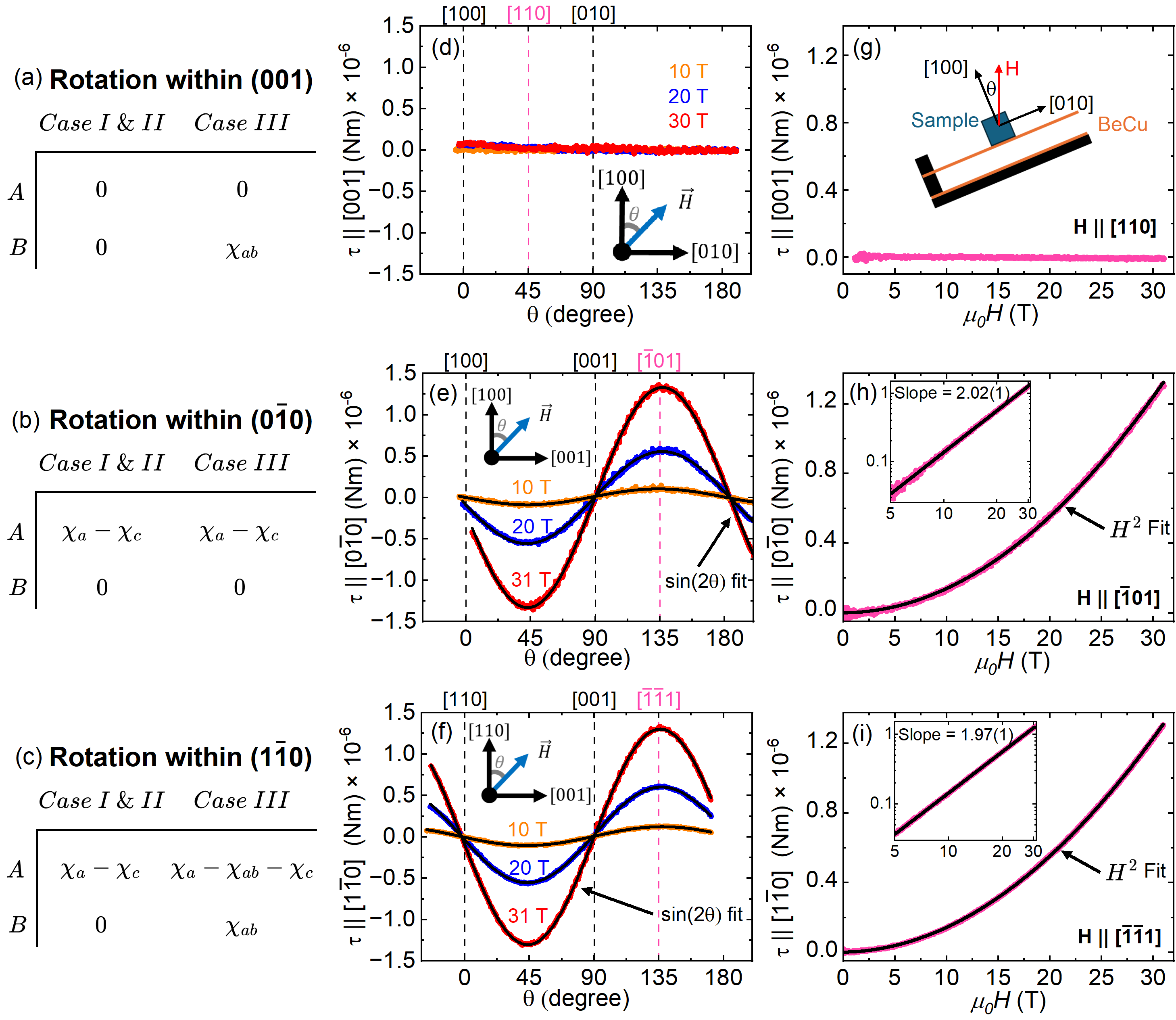}
    \caption{Torque magnetometry on a RuO$_2$ single crystal at 40 K. (a)-(c) Torque expressions for Cases I-III (Fig.~\ref{intro}) for the three measurement configurations described in the text. $A$ and $B$ are coefficients in $\tau(\theta) =\mu_0^2 H^2[\frac{A}{2}\sin(2\theta)+B(2\sin^2(\theta)-1)]$. (d)-(f) Angular-dependent torque for rotations of the magnetic field within the (001), $(0\overline{1}0)$, and $(1\overline{1}0)$ planes. Data for the (d) (001) configuration are not periodic, whereas data for the (e) $(0\overline{1}0)$ and (f) $(1\overline{1}0)$ configurations are described by a $\sin(2\theta)$ form. (g)-(i) Field-dependent torque amplitude for fixed directions within the (001), $(0\overline{1}0)$, $(1\overline{1}0)$ planes. Data for the (g) (001) configuration do not have a measurable field dependence up to 31 T, whereas data for the (h) $(0\overline{1}0)$ and (i) $(1\overline{1}0)$ configurations scale quadratically with magnetic field. Insets of (h) and (i): Log-Log fitting of torque amplitude showing quadratic scaling. A schematic of the experiment is shown in (g).}
    \label{ani_torque}
\end{figure*}

It follows that the experimental determination of the magnetic susceptibility tensor will constrain the type of magnetic order in RuO$_2$. Toward this end, we use a direct, sensitive, and thermodynamic probe of small anisotropies in magnetic susceptibility: torque magnetometry~\cite{perfetti2017cantilever}. Magnetic torque measurements consist of rotating the magnetic field within a crystallographic plane and measuring the torque normal to that plane (Fig.~\ref{ani_torque}). As shown in the SM~\cite{SM}, an easy axis in a direction other than the $c-$axis would, at a minimum, cause a finite torque for magnetic fields rotated within the (001) plane and differences in the torque response for fields within the (0$\bar{1}$0) and (1$\bar{1}$0) planes (see expressions in Fig.~\ref{ani_torque}). Magnetic domains would also affect the magnetic anisotropy by changing the periodicity of the torque ($\it{e.g.}$, torque measurements on MnTe with domains \cite{komatsubara1963magnetic}). Thus, experimental configurations were adopted to rotate the magnetic field within the (001), (0$\bar{1}$0), and (1$\bar{1}$0) planes.

{}

\begin{figure}[t]
    \centering
    \includegraphics[width=2.7in, clip]{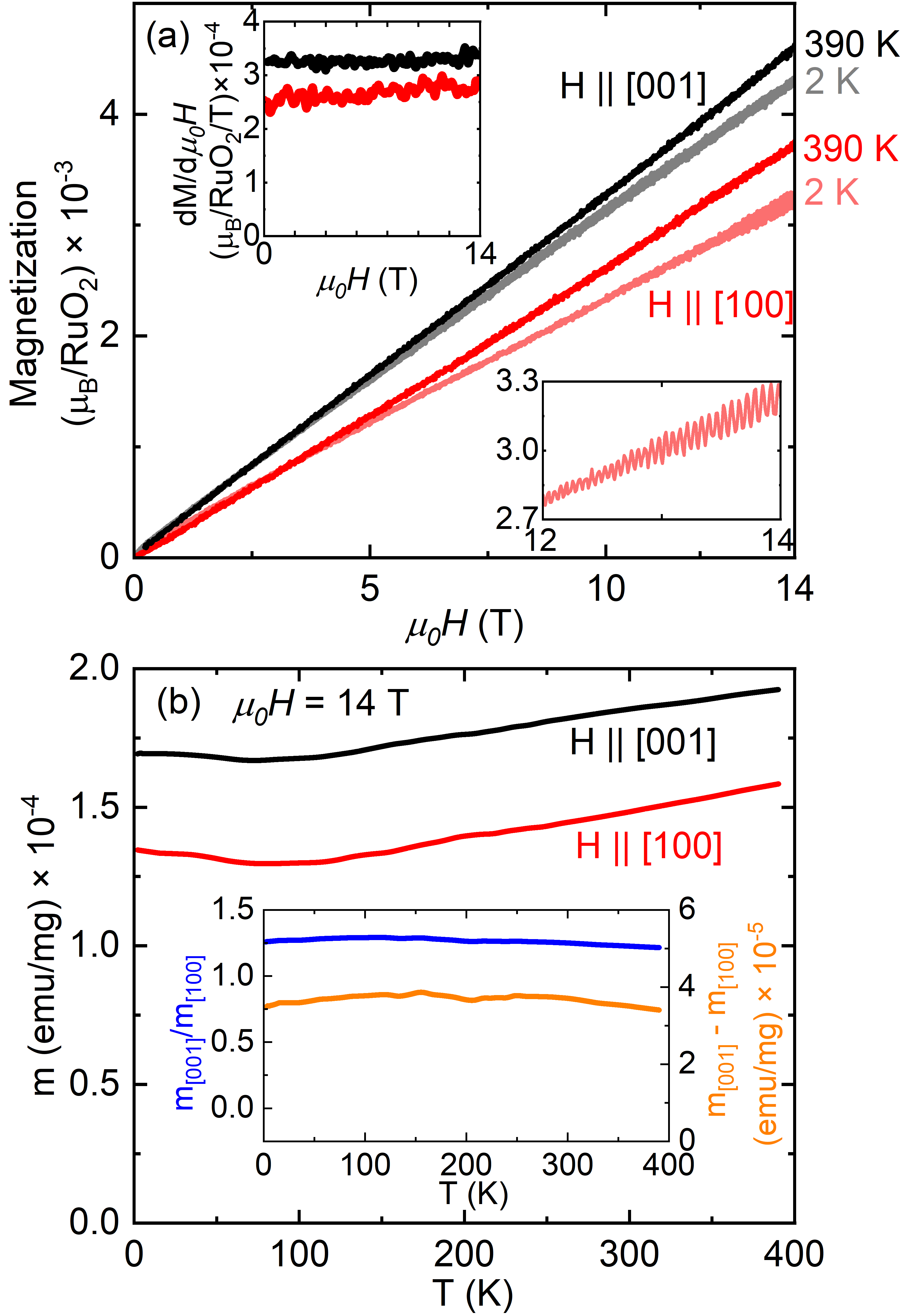}
    \caption{(a) Longitudinal magnetization for different directions and temperatures. Quantum oscillations are present at high fields and low temperatures. Left inset: differential susceptibility demonstrating linear-in-field magnetization. Right inset: magnetic quantum oscillations when the field is along [100] at 2 K. (b) Temperature-dependent moment along different directions. Inset: temperature dependence of m$_{[001]}$/m$_{[100]}$ and m$_{[001]}$-m$_{[100]}$ to quantify anisotropy.}
     \label{ani_VSM}
\end{figure}

\begin{figure}[b]
    \centering
    \includegraphics[width=3.5in]{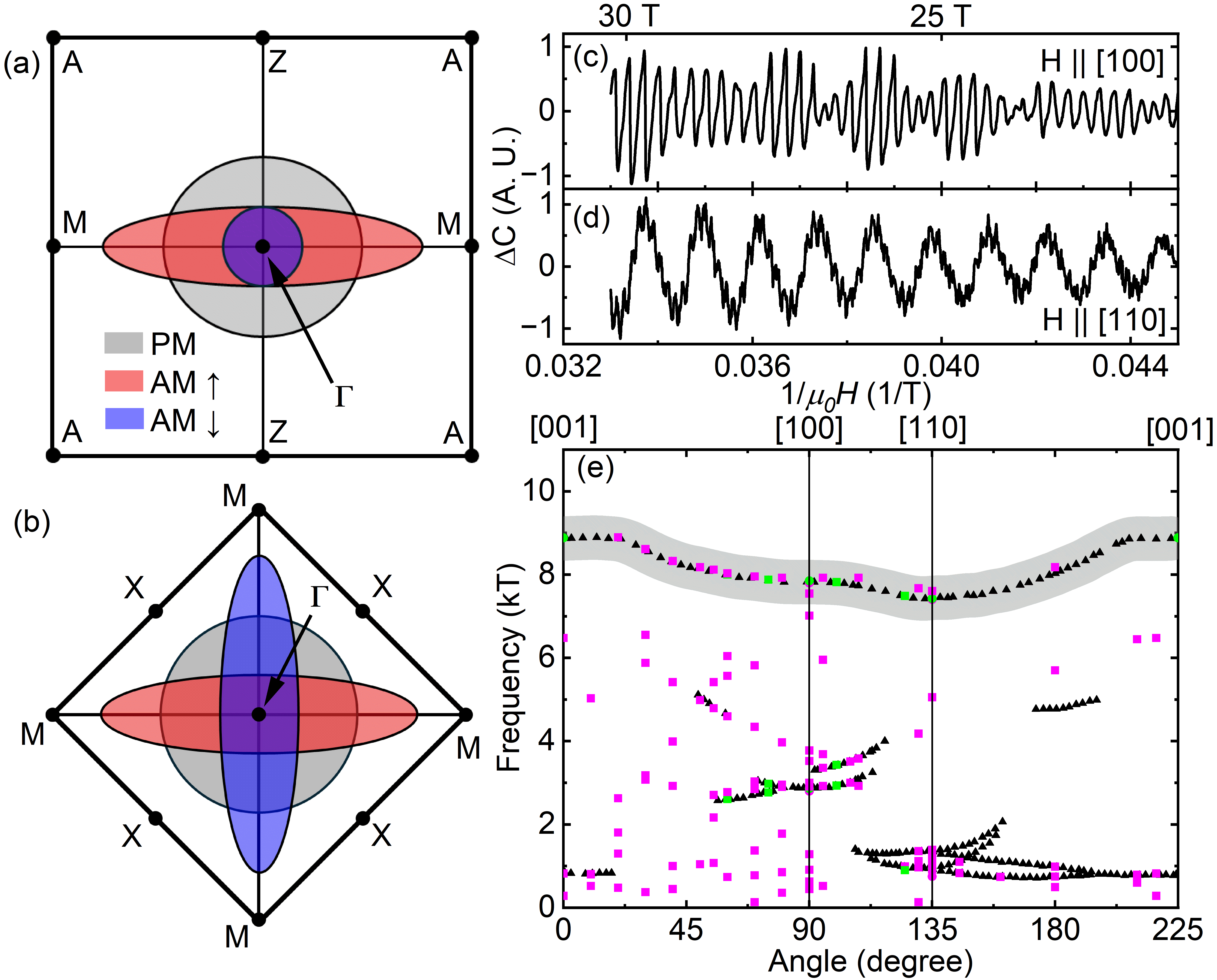}
    \caption{Fermi surface of RuO$_2$. (a),(b) Sketches of the Fermi surface pocket at the $\Gamma$ point assuming paramagnetism and altermagnetism along the (a) $\Gamma$M and (b) $\Gamma$Z directions. The pocket in the paramagnetic case is shown in gray. Blue and red depict the spin-split pocket in the presence of altermagnetic order. (c),(d) Examples of quantum oscillations measured with torque magnetometry at 0.5 K in two directions after polynomial subtraction. (e) Quantum oscillation frequencies with respect to magnetic field direction. Symbols denote different techniques (circles are magnetization, squares are torque magnetometry) and colors indicate different samples. Triangles are from prior work \cite{graebner1976magnetothermal}. The highest frequencies, corresponding to a nearly spherical pocket around the $\Gamma$ point, are shaded in gray.}
    \label{QO}
\end{figure}

Figures~\ref{ani_torque}(d,g) demonstrate that rotating the magnetic field within the (001) plane produces no discernible torque up to 31 T (see SM for details~\cite{SM}). In contrast, rotating the magnetic field within the (0$\bar{1}$0) and (1$\bar{1}$0) planes yields identical torque responses up to 31 T: for a fixed magnetic field, the torque in both configurations has the same sinusoidal dependence on angle with an equivalent amplitude and periodicity of 180\degree. Furthermore, Figs.~\ref{ani_torque}(h,i) show the torque amplitudes do not deviate from field squared scaling within these two planes. Note, data in Fig.~\ref{ani_torque} were taken at a single temperature (40 K) because the magnetic susceptibility magnitudes and anisotropy are weakly temperature dependent (Fig.~\ref{ani_VSM}), and the effects of quantum oscillations are minimized (Fig.~\ref{QO}).

The three angular and field dependencies in Fig.~\ref{ani_torque} are only simultaneously satisfied if RuO$_2$ single crystals have uniaxial magnetic anisotropy with isotropic susceptibility within the $ab-$plane. Thus, magnetic orders with a component of the N\'eel vector not parallel to the $c-$axis, including Case III in Fig.~\ref{intro}, are incompatible with the data in Fig.~\ref{ani_torque}. These measurements also rule out magnetic domains and field-induced reorientation of a N\'eel vector, as has been reported in thin film RuO$_2$ \cite{feng2022anomalous} and other rutile magnets \cite{Bazan1975weak}.

To distinguish between Cases I and II, quantitative arguments which move beyond symmetry are necessary. It is well-established that systems with local or itinerant collinear magnetic order exhibit a smaller magnetic susceptibility parallel to the N\'eel vector that approaches zero at 0 K and a larger, temperature-independent magnetic susceptibility perpendicular to the N\'eel vector for small magnetic fields \cite{blundell2001magnetism, Lidiard1953}. Therefore, one would expect $\chi_a > \chi_c$ with $\lim_{T\to0}\chi_c = 0$ in Case II.

The first piece of evidence supporting paramagnetic Case I over altermagnetic Case II is the sign of the torque, which indicates $\chi_c > \chi_{a}$ (Fig.~\ref{ani_torque}). To corroborate $\chi_c > \chi_{a}$ and check if $\lim_{T\to0}\chi_c = 0$, we measure the longitudinal magnetization along the $a$- and $c$-axes (Fig.~\ref{ani_VSM}). Both directions are dominated by linear-in-field magnetization at all temperatures (see SM~\cite{SM}). The magnitudes of the magnetization along these two directions show a similar, weak temperature dependence such that the magnetic anisotropy minimally varies with temperature, in agreement with previous reports~\cite{ryden1970magnetic,Kiefer2025}. Consequently, the magnetization for a given magnetic field remains approximately 20\% larger along the $c-$axis at all temperatures, and the magnetic susceptibility does not tend toward zero in the limit of 0 K in either principal crystallographic direction.

{}

Therefore, the magnetization data in Fig.~\ref{ani_VSM} reinforce the torque evidence for paramagnetic Case I. In fact, the magnetic anisotropy $\chi_{a}-\chi_c$ determined from magnetization ($2.4(2)\times10^{-5}$) and torque ($2.3(1)\times10^{-5}$) are in quantitative agreement \cite{SM}. It is also worth noting our RuO$_2$ single crystal results quantitatively agree with other measurements on single crystals~\cite{ryden1970magnetic,Kiefer2025}, but substantially differ from some RuO$_2$ thin film measurements \cite{feng2022anomalous,tschirner2023saturation}. First, we observe no phase transitions from 0.5-400 K and up to 31 T (thin films show a zero-field transition near 380 K~\cite{feng2022anomalous}). Second, the susceptibilities in Fig.~\ref{ani_VSM} are two orders of magnitude smaller than those reported for thin films~\cite{feng2022anomalous}. Third, Fig.~\ref{ani_torque} shows no evidence for spin-reorientation when a component of the magnetic field is along [110] up to 31 T, whereas this occurs $\sim$1 T in thin films~\cite{feng2022anomalous}.

{}

An additional stringent test for paramagnetic RuO$_2$ single crystals is characterizing the Fermi surface of RuO$_2$ because the magnetic properties of itinerant paramagnets reflect their Fermi surface. Furthermore, density functional theory calculations indicate pronounced differences between paramagnetic and altermagnetic Fermi surfaces, particularly in the shape of the Brillouin zone center pocket: this pocket is nearly spherical for paramagnetic RuO$_2$, whereas it is flatter with a four-fold distortion reflecting $d$-wave altermagnetism in altermagnetic RuO$_2$ (Fig.~\ref{QO} (a, b)). Table~\ref{table:DFT} quantifies the predicted differences in the Fermi surface shape via ratios of the Fermi surface radii along the $\Gamma M$, $\Gamma Z$, and $\Gamma X$ paths. Although there is some scatter in the altermagnetic calculations, altermagnetic Fermi surfaces are significantly more anisotropic than paramagnetic Fermi surfaces.

We directly characterized the Fermi surface with quantum oscillations~\cite{Shoenberg} measured by torque magnetometry (Fig.~\ref{QO}(c, d)) and magnetization (SM~\cite{SM}) to determine the shape of the $\Gamma$ point pocket and distinguish between a paramagnetic or altermagnetic Fermi surface. The principal features of the angular-dependent quantum oscillation frequencies in Fig. \ref{QO} (e) are a set of high frequencies that vary minimally with angle (7.4-8.9 kT) and a range of lower frequencies with complex angular dependencies (see SM for details \cite{SM}). The former are associated with a nearly-spherical Fermi surface pocket at the Brillouin zone center, and the latter are attributed to smaller, non-spherical pocket(s) \cite{graebner1976magnetothermal, ahn2019antiferromagnetism, huang2024ab}. These features agree well with previous measurements of quantum oscillations in RuO$_2$ single crystals~\cite{graebner1976magnetothermal, marcus1968measurement}, however we observed additional lower frequencies owing to the availability of higher fields and lower temperatures. Crucially, these measurements reveal a nearly spherical $\Gamma$ point pocket (Fig.~\ref{QO}) which is in quantitative agreement with theoretical paramagnetic Fermi surfaces (Table~\ref{table:DFT}). Thus, the experimental Fermi surface of RuO$_2$ is consistent with paramagnetism, not altermagnetism.

\begin{table}

\renewcommand{\arraystretch}{1.5}
\caption{Fermi surface anisotropy at the $\Gamma$ point from our measurements, literature  measurements, and literature DFT calculations assuming paramagnetism (Case I) and altermagnetism (Case II). Values are the ratios $\Gamma$M/$\Gamma$Z and $\Gamma$X/$\Gamma$M of the $\Gamma$ point pocket. }
\centering
\begin{tabular}{cccc}
\hline
& Quantum Oscillations & Paramagnetic & Altermagnetic          \\
& (Exp.) & (DFT) & (DFT)          \\
\hline
 & 1.13(1) [This work] &  1.10 \cite{huang2024ab}  & 3.25\cite{huang2024ab}     \\
 $\Gamma$M/$\Gamma$Z & 1.23 \cite{graebner1976magnetothermal} & 0.91 \cite{ahn2019antiferromagnetism}& 1.53 \cite{ahn2019antiferromagnetism}        \\
 &  &   & 3.87\cite{samanta2024tunneling}       \\
\hline
 & 1.06(1) [This work] &  1.05 \cite{huang2024ab}  & 0.36 \cite{huang2024ab}      \\
 & 1.00 \cite{graebner1976magnetothermal} & 1.03 \cite{ahn2019antiferromagnetism}& 1.00 \cite{ahn2019antiferromagnetism}       \\
 $\Gamma$X/$\Gamma$M &   & 1.10 \cite{fedchenko2024observation}   & 0.71 \cite{fedchenko2024observation}      \\
  &   &  & 0.44 \cite{smejkal2020}       \\
  &   &  & 0.2\cite{samanta2024tunneling}        \\
\hline
\end{tabular}
\label{table:DFT}
\end{table}

Prior to recent measurements~\cite{berlijn2017itinerant} and work on altermagnetism~\cite{smejkal2020}, the prevailing interpretation was RuO$_2$ is an itinerant paramagnet. For example, the Pauli paramagnetic susceptibility of RuO$_2$ determined from the Sommerfeld coefficient \cite{glassford1993electronic} semi-quantitatively agrees with the measured magnetic susceptibility \cite{ryden1970magnetic}. Our results are consistent with this interpretation and indicate high-quality, single crystals of RuO$_2$ are itinerant paramagnets. We reproduced our findings on multiple samples, instruments, and techniques (see Fig.~\ref{QO} and SM \cite{SM}) and confirmed sample quality by determining the electron mean free path ($\approx$ 160 unit cells) and residual-resisitivity ratio ($\approx 250$). In addition to the symmetry of the magnetic susceptibility tensor, relative size of the magnetic susceptibility components, and shape of the Fermi surface, other observations in Figs.~\ref{ani_torque} and \ref{ani_VSM} are also consistent with itinerant paramagnetism: (1) $d\chi/dT > 0$ implies a local minimum in the electronic density of states at the Fermi level~\cite{Stoner1936,Kriessman1954}, in agreement with other RuO$_2$ work~\cite{cox1986electronic, glassford1993electronic}; (2) the small, low temperature upturns in $\chi_a$ and $\chi_c$ reflect contributions from multiple bands at the Fermi level \cite{shimizu1960magnetic,Mattheiss1976,cox1986electronic,pawula2024multiband}; and (3) the small magnitudes and weak temperature dependencies of $\chi_a$ and $\chi_c$ and nearly temperature-independent anisotropy are consequences of magnetization originating from spin-polarization of the Fermi surface~\cite{Galoshina1974, Shimizu1981}.

So what of altermagnetism in RuO$_2$? Our measurements indicate high-quality RuO$_2$ single crystals do not exhibit long-range magnetic order, but there may be extrinsic avenues to engineer altermagnetism in RuO$_2$. For example, recent thin film experiments have shown the importance of epitaxial strain and thickness in inducing magnetic order in RuO$_2$ \cite{jeong2025metallicity} and calculations suggest point defects can drive RuO$_2$ toward a magnetic state~\cite{smolyanyuk2024fragility}. Such possibilities warrant further study.

In summary, we have used a combination of thermodynamic probes to characterize magnetic order in RuO$_2$ single crystals. Our measurements indicate uniaxial anisotropy with isotropic magnetic susceptibility within the $ab-$plane and $\chi_c > \chi_a$ up to 31 T in all investigated temperatures. Magnetic quantum oscillations demonstrate a nearly-spherical Fermi surface pocket around the Brillouin zone center, in agreement with the theoretical paramagnetic Fermi surface. Together, these results strongly suggest that RuO$_2$ single crystals are itinerant paramagnets without long-range magnetic order and, accordingly, not altermagnets.

\medskip

\section*{Acknowledgments}
\noindent T.Q., B.M., M.L., and C.A.M. acknowledge the support of the Laboratory Directed Research and Development program of Los Alamos National Laboratory under project number 20240225ER. A.R. and H.Z. acknowledge support from the U.S. Department of Energy with Grant No.DE-SC0020254. A portion of this work was performed at the National High Magnetic Field Laboratory, which is supported by the National Science Foundation Cooperative Agreement No. DMR-2128556, the State of Florida, and the Department of Energy. The authors thank Shermane Benjamin and Clyde Martin for technical assistance at the National High Magnetic Field Laboratory's DC Field Facility.

\medskip

\bibliographystyle{apsrev4-1}
\bibliography{RuO2}

\begin{thebibliography}{71}%
\makeatletter
\providecommand \@ifxundefined [1]{%
 \@ifx{#1\undefined}
}%
\providecommand \@ifnum [1]{%
 \ifnum #1\expandafter \@firstoftwo
 \else \expandafter \@secondoftwo
 \fi
}%
\providecommand \@ifx [1]{%
 \ifx #1\expandafter \@firstoftwo
 \else \expandafter \@secondoftwo
 \fi
}%
\providecommand \natexlab [1]{#1}%
\providecommand \enquote  [1]{``#1''}%
\providecommand \bibnamefont  [1]{#1}%
\providecommand \bibfnamefont [1]{#1}%
\providecommand \citenamefont [1]{#1}%
\providecommand \href@noop [0]{\@secondoftwo}%
\providecommand \href [0]{\begingroup \@sanitize@url \@href}%
\providecommand \@href[1]{\@@startlink{#1}\@@href}%
\providecommand \@@href[1]{\endgroup#1\@@endlink}%
\providecommand \@sanitize@url [0]{\catcode `\\12\catcode `\$12\catcode `\&12\catcode `\#12\catcode `\^12\catcode `\_12\catcode `\%12\relax}%
\providecommand \@@startlink[1]{}%
\providecommand \@@endlink[0]{}%
\providecommand \url  [0]{\begingroup\@sanitize@url \@url }%
\providecommand \@url [1]{\endgroup\@href {#1}{\urlprefix }}%
\providecommand \urlprefix  [0]{URL }%
\providecommand \Eprint [0]{\href }%
\providecommand \doibase [0]{http://dx.doi.org/}%
\providecommand \selectlanguage [0]{\@gobble}%
\providecommand \bibinfo  [0]{\@secondoftwo}%
\providecommand \bibfield  [0]{\@secondoftwo}%
\providecommand \translation [1]{[#1]}%
\providecommand \BibitemOpen [0]{}%
\providecommand \bibitemStop [0]{}%
\providecommand \bibitemNoStop [0]{.\EOS\space}%
\providecommand \EOS [0]{\spacefactor3000\relax}%
\providecommand \BibitemShut  [1]{\csname bibitem#1\endcsname}%
\let\auto@bib@innerbib\@empty
\bibitem [{\citenamefont {Borovik-Romanov}\ \emph {et~al.}(2013)\citenamefont {Borovik-Romanov}, \citenamefont {Grimmer},\ and\ \citenamefont {Kenzelmann}}]{BorovikRomanov2013}%
  \BibitemOpen
  \bibfield  {author} {\bibinfo {author} {\bibfnamefont {A.~S.}\ \bibnamefont {Borovik-Romanov}}, \bibinfo {author} {\bibfnamefont {H.}~\bibnamefont {Grimmer}}, \ and\ \bibinfo {author} {\bibfnamefont {M.}~\bibnamefont {Kenzelmann}},\ }\enquote {\bibinfo {title} {Magnetic properties},}\ in\ \href@noop {} {\emph {\bibinfo {booktitle} {International Tables for Crystallography}}}\ (\bibinfo  {publisher} {John Wiley \& Sons, Ltd},\ \bibinfo {year} {2013})\ Chap.\ \bibinfo {chapter} {1.5}, pp.\ \bibinfo {pages} {106--152}\BibitemShut {NoStop}%
\bibitem [{\citenamefont {Hirohata}\ \emph {et~al.}(2020)\citenamefont {Hirohata}, \citenamefont {Yamada}, \citenamefont {Nakatani}, \citenamefont {Prejbeanu}, \citenamefont {Diény}, \citenamefont {Pirro},\ and\ \citenamefont {Hillebrands}}]{Hirohata2020}%
  \BibitemOpen
  \bibfield  {author} {\bibinfo {author} {\bibfnamefont {A.}~\bibnamefont {Hirohata}}, \bibinfo {author} {\bibfnamefont {K.}~\bibnamefont {Yamada}}, \bibinfo {author} {\bibfnamefont {Y.}~\bibnamefont {Nakatani}}, \bibinfo {author} {\bibfnamefont {I.-L.}\ \bibnamefont {Prejbeanu}}, \bibinfo {author} {\bibfnamefont {B.}~\bibnamefont {Diény}}, \bibinfo {author} {\bibfnamefont {P.}~\bibnamefont {Pirro}}, \ and\ \bibinfo {author} {\bibfnamefont {B.}~\bibnamefont {Hillebrands}},\ }\href@noop {} {\bibfield  {journal} {\bibinfo  {journal} {Journal of Magnetism and Magnetic Materials}\ }\textbf {\bibinfo {volume} {509}},\ \bibinfo {pages} {166711} (\bibinfo {year} {2020})}\BibitemShut {NoStop}%
\bibitem [{\citenamefont {Jungwirth}\ and\ \citenamefont {Wunderlich}(2016)}]{Jungwirth2016}%
  \BibitemOpen
  \bibfield  {author} {\bibinfo {author} {\bibfnamefont {M.~X. W.~P.}\ \bibnamefont {Jungwirth}, \bibfnamefont {T.}}\ and\ \bibinfo {author} {\bibfnamefont {J.}~\bibnamefont {Wunderlich}},\ }\href@noop {} {\bibfield  {journal} {\bibinfo  {journal} {Nature Nanotechnology}\ }\textbf {\bibinfo {volume} {11}},\ \bibinfo {pages} {231} (\bibinfo {year} {2016})}\BibitemShut {NoStop}%
\bibitem [{\citenamefont {{\v{S}}mejkal}\ \emph {et~al.}(2022{\natexlab{a}})\citenamefont {{\v{S}}mejkal}, \citenamefont {Sinova},\ and\ \citenamefont {Jungwirth}}]{vsmejkal2022beyond}%
  \BibitemOpen
  \bibfield  {author} {\bibinfo {author} {\bibfnamefont {L.}~\bibnamefont {{\v{S}}mejkal}}, \bibinfo {author} {\bibfnamefont {J.}~\bibnamefont {Sinova}}, \ and\ \bibinfo {author} {\bibfnamefont {T.}~\bibnamefont {Jungwirth}},\ }\href@noop {} {\bibfield  {journal} {\bibinfo  {journal} {Physical Review X}\ }\textbf {\bibinfo {volume} {12}},\ \bibinfo {pages} {031042} (\bibinfo {year} {2022}{\natexlab{a}})}\BibitemShut {NoStop}%
\bibitem [{\citenamefont {{\v{S}}mejkal}\ \emph {et~al.}(2022{\natexlab{b}})\citenamefont {{\v{S}}mejkal}, \citenamefont {Sinova},\ and\ \citenamefont {Jungwirth}}]{vsmejkal2022emerging}%
  \BibitemOpen
  \bibfield  {author} {\bibinfo {author} {\bibfnamefont {L.}~\bibnamefont {{\v{S}}mejkal}}, \bibinfo {author} {\bibfnamefont {J.}~\bibnamefont {Sinova}}, \ and\ \bibinfo {author} {\bibfnamefont {T.}~\bibnamefont {Jungwirth}},\ }\href@noop {} {\bibfield  {journal} {\bibinfo  {journal} {Physical Review X}\ }\textbf {\bibinfo {volume} {12}},\ \bibinfo {pages} {040501} (\bibinfo {year} {2022}{\natexlab{b}})}\BibitemShut {NoStop}%
\bibitem [{\citenamefont {Litvin}\ and\ \citenamefont {Opechowski}(1974)}]{Litvin1974}%
  \BibitemOpen
  \bibfield  {author} {\bibinfo {author} {\bibfnamefont {D.}~\bibnamefont {Litvin}}\ and\ \bibinfo {author} {\bibfnamefont {W.}~\bibnamefont {Opechowski}},\ }\href@noop {} {\bibfield  {journal} {\bibinfo  {journal} {Physica}\ }\textbf {\bibinfo {volume} {76}},\ \bibinfo {pages} {538} (\bibinfo {year} {1974})}\BibitemShut {NoStop}%
\bibitem [{\citenamefont {Bhowal}\ and\ \citenamefont {Spaldin}(2024)}]{Bhowal2024}%
  \BibitemOpen
  \bibfield  {author} {\bibinfo {author} {\bibfnamefont {S.}~\bibnamefont {Bhowal}}\ and\ \bibinfo {author} {\bibfnamefont {N.~A.}\ \bibnamefont {Spaldin}},\ }\href {\doibase 10.1103/PhysRevX.14.011019} {\bibfield  {journal} {\bibinfo  {journal} {Physical Review X}\ }\textbf {\bibinfo {volume} {14}},\ \bibinfo {pages} {011019} (\bibinfo {year} {2024})}\BibitemShut {NoStop}%
\bibitem [{\citenamefont {McClarty}\ and\ \citenamefont {Rau}(2024)}]{McClarty2024}%
  \BibitemOpen
  \bibfield  {author} {\bibinfo {author} {\bibfnamefont {P.~A.}\ \bibnamefont {McClarty}}\ and\ \bibinfo {author} {\bibfnamefont {J.~G.}\ \bibnamefont {Rau}},\ }\href {\doibase 10.1103/PhysRevLett.132.176702} {\bibfield  {journal} {\bibinfo  {journal} {Physical Review Letters}\ }\textbf {\bibinfo {volume} {132}},\ \bibinfo {pages} {176702} (\bibinfo {year} {2024})}\BibitemShut {NoStop}%
\bibitem [{\citenamefont {Hayami}\ \emph {et~al.}(2019)\citenamefont {Hayami}, \citenamefont {Yanagi},\ and\ \citenamefont {Kusunose}}]{Hayami2019}%
  \BibitemOpen
  \bibfield  {author} {\bibinfo {author} {\bibfnamefont {S.}~\bibnamefont {Hayami}}, \bibinfo {author} {\bibfnamefont {Y.}~\bibnamefont {Yanagi}}, \ and\ \bibinfo {author} {\bibfnamefont {H.}~\bibnamefont {Kusunose}},\ }\href@noop {} {\bibfield  {journal} {\bibinfo  {journal} {Journal of the Physical Society of Japan}\ }\textbf {\bibinfo {volume} {88}},\ \bibinfo {pages} {123702} (\bibinfo {year} {2019})}\BibitemShut {NoStop}%
\bibitem [{\citenamefont {Yuan}\ \emph {et~al.}(2020)\citenamefont {Yuan}, \citenamefont {Wang}, \citenamefont {Luo}, \citenamefont {Rashba},\ and\ \citenamefont {Zunger}}]{Yuan2020}%
  \BibitemOpen
  \bibfield  {author} {\bibinfo {author} {\bibfnamefont {L.-D.}\ \bibnamefont {Yuan}}, \bibinfo {author} {\bibfnamefont {Z.}~\bibnamefont {Wang}}, \bibinfo {author} {\bibfnamefont {J.-W.}\ \bibnamefont {Luo}}, \bibinfo {author} {\bibfnamefont {E.~I.}\ \bibnamefont {Rashba}}, \ and\ \bibinfo {author} {\bibfnamefont {A.}~\bibnamefont {Zunger}},\ }\href {\doibase 10.1103/PhysRevB.102.014422} {\bibfield  {journal} {\bibinfo  {journal} {Physical Review B}\ }\textbf {\bibinfo {volume} {102}},\ \bibinfo {pages} {014422} (\bibinfo {year} {2020})}\BibitemShut {NoStop}%
\bibitem [{\citenamefont {Jungwirth}\ \emph {et~al.}(2024)\citenamefont {Jungwirth}, \citenamefont {Fernandes}, \citenamefont {Sinova},\ and\ \citenamefont {Smejkal}}]{jungwirth2024}%
  \BibitemOpen
  \bibfield  {author} {\bibinfo {author} {\bibfnamefont {T.}~\bibnamefont {Jungwirth}}, \bibinfo {author} {\bibfnamefont {R.~M.}\ \bibnamefont {Fernandes}}, \bibinfo {author} {\bibfnamefont {J.}~\bibnamefont {Sinova}}, \ and\ \bibinfo {author} {\bibfnamefont {L.}~\bibnamefont {Smejkal}},\ }\href {https://arxiv.org/abs/2409.10034} {\enquote {\bibinfo {title} {Altermagnets and beyond: Nodal magnetically-ordered phases},}\ } (\bibinfo {year} {2024}),\ \Eprint {http://arxiv.org/abs/2409.10034} {arXiv:2409.10034 [cond-mat.mtrl-sci]} \BibitemShut {NoStop}%
\bibitem [{\citenamefont {Jungwirth}\ \emph {et~al.}(2025)\citenamefont {Jungwirth}, \citenamefont {Fernandes}, \citenamefont {Fradkin}, \citenamefont {MacDonald}, \citenamefont {Sinova},\ and\ \citenamefont {Smejkal}}]{jungwirth2025}%
  \BibitemOpen
  \bibfield  {author} {\bibinfo {author} {\bibfnamefont {T.}~\bibnamefont {Jungwirth}}, \bibinfo {author} {\bibfnamefont {R.~M.}\ \bibnamefont {Fernandes}}, \bibinfo {author} {\bibfnamefont {E.}~\bibnamefont {Fradkin}}, \bibinfo {author} {\bibfnamefont {A.~H.}\ \bibnamefont {MacDonald}}, \bibinfo {author} {\bibfnamefont {J.}~\bibnamefont {Sinova}}, \ and\ \bibinfo {author} {\bibfnamefont {L.}~\bibnamefont {Smejkal}},\ }\href {https://arxiv.org/abs/2411.00717} {\enquote {\bibinfo {title} {Altermagnetism: an unconventional spin-ordered phase of matter},}\ } (\bibinfo {year} {2025}),\ \Eprint {http://arxiv.org/abs/2411.00717} {arXiv:2411.00717 [cond-mat.mtrl-sci]} \BibitemShut {NoStop}%
\bibitem [{\citenamefont {Hellenes}\ \emph {et~al.}(2024)\citenamefont {Hellenes}, \citenamefont {Jungwirth}, \citenamefont {Jaeschke-Ubiergo}, \citenamefont {Chakraborty}, \citenamefont {Sinova},\ and\ \citenamefont {Šmejkal}}]{hellenes2024pwavemagnets}%
  \BibitemOpen
  \bibfield  {author} {\bibinfo {author} {\bibfnamefont {A.~B.}\ \bibnamefont {Hellenes}}, \bibinfo {author} {\bibfnamefont {T.}~\bibnamefont {Jungwirth}}, \bibinfo {author} {\bibfnamefont {R.}~\bibnamefont {Jaeschke-Ubiergo}}, \bibinfo {author} {\bibfnamefont {A.}~\bibnamefont {Chakraborty}}, \bibinfo {author} {\bibfnamefont {J.}~\bibnamefont {Sinova}}, \ and\ \bibinfo {author} {\bibfnamefont {L.}~\bibnamefont {Šmejkal}},\ }\href {https://arxiv.org/abs/2309.01607} {\enquote {\bibinfo {title} {P-wave magnets},}\ } (\bibinfo {year} {2024}),\ \Eprint {http://arxiv.org/abs/2309.01607} {arXiv:2309.01607 [cond-mat.mes-hall]} \BibitemShut {NoStop}%
\bibitem [{\citenamefont {Cheong}\ and\ \citenamefont {Huang}(2024)}]{cheong2024altermagnetism}%
  \BibitemOpen
  \bibfield  {author} {\bibinfo {author} {\bibfnamefont {S.-W.}\ \bibnamefont {Cheong}}\ and\ \bibinfo {author} {\bibfnamefont {F.-T.}\ \bibnamefont {Huang}},\ }\href@noop {} {\bibfield  {journal} {\bibinfo  {journal} {npj Quantum Materials}\ }\textbf {\bibinfo {volume} {9}},\ \bibinfo {pages} {13} (\bibinfo {year} {2024})}\BibitemShut {NoStop}%
\bibitem [{\citenamefont {Yuan}\ \emph {et~al.}(2024)\citenamefont {Yuan}, \citenamefont {Georgescu},\ and\ \citenamefont {Rondinelli}}]{yuan2024nonrelativistic}%
  \BibitemOpen
  \bibfield  {author} {\bibinfo {author} {\bibfnamefont {L.-D.}\ \bibnamefont {Yuan}}, \bibinfo {author} {\bibfnamefont {A.~B.}\ \bibnamefont {Georgescu}}, \ and\ \bibinfo {author} {\bibfnamefont {J.~M.}\ \bibnamefont {Rondinelli}},\ }\href@noop {} {\bibfield  {journal} {\bibinfo  {journal} {Physical Review Letters}\ }\textbf {\bibinfo {volume} {133}},\ \bibinfo {pages} {216701} (\bibinfo {year} {2024})}\BibitemShut {NoStop}%
\bibitem [{\citenamefont {Xu}\ \emph {et~al.}(2025)\citenamefont {Xu}, \citenamefont {Wu}, \citenamefont {Zhi}, \citenamefont {Cao}, \citenamefont {Dai}, \citenamefont {Cao}, \citenamefont {Wang},\ and\ \citenamefont {Lin}}]{Xu_2025}%
  \BibitemOpen
  \bibfield  {author} {\bibinfo {author} {\bibfnamefont {C.}~\bibnamefont {Xu}}, \bibinfo {author} {\bibfnamefont {S.}~\bibnamefont {Wu}}, \bibinfo {author} {\bibfnamefont {G.-X.}\ \bibnamefont {Zhi}}, \bibinfo {author} {\bibfnamefont {G.}~\bibnamefont {Cao}}, \bibinfo {author} {\bibfnamefont {J.}~\bibnamefont {Dai}}, \bibinfo {author} {\bibfnamefont {C.}~\bibnamefont {Cao}}, \bibinfo {author} {\bibfnamefont {X.}~\bibnamefont {Wang}}, \ and\ \bibinfo {author} {\bibfnamefont {H.-Q.}\ \bibnamefont {Lin}},\ }\href {\doibase 10.1038/s41467-025-58446-6} {\bibfield  {journal} {\bibinfo  {journal} {Nature Communications}\ }\textbf {\bibinfo {volume} {16}} (\bibinfo {year} {2025}),\ 10.1038/s41467-025-58446-6}\BibitemShut {NoStop}%
\bibitem [{\citenamefont {Reimers}\ \emph {et~al.}(2024)\citenamefont {Reimers}, \citenamefont {Odenbreit}, \citenamefont {{\v{S}}mejkal}, \citenamefont {Strocov}, \citenamefont {Constantinou}, \citenamefont {Hellenes}, \citenamefont {Jaeschke~Ubiergo}, \citenamefont {Campos}, \citenamefont {Bharadwaj}, \citenamefont {Chakraborty} \emph {et~al.}}]{reimers2024direct}%
  \BibitemOpen
  \bibfield  {author} {\bibinfo {author} {\bibfnamefont {S.}~\bibnamefont {Reimers}}, \bibinfo {author} {\bibfnamefont {L.}~\bibnamefont {Odenbreit}}, \bibinfo {author} {\bibfnamefont {L.}~\bibnamefont {{\v{S}}mejkal}}, \bibinfo {author} {\bibfnamefont {V.~N.}\ \bibnamefont {Strocov}}, \bibinfo {author} {\bibfnamefont {P.}~\bibnamefont {Constantinou}}, \bibinfo {author} {\bibfnamefont {A.~B.}\ \bibnamefont {Hellenes}}, \bibinfo {author} {\bibfnamefont {R.}~\bibnamefont {Jaeschke~Ubiergo}}, \bibinfo {author} {\bibfnamefont {W.~H.}\ \bibnamefont {Campos}}, \bibinfo {author} {\bibfnamefont {V.~K.}\ \bibnamefont {Bharadwaj}}, \bibinfo {author} {\bibfnamefont {A.}~\bibnamefont {Chakraborty}},  \emph {et~al.},\ }\href@noop {} {\bibfield  {journal} {\bibinfo  {journal} {Nature Communications}\ }\textbf {\bibinfo {volume} {15}},\ \bibinfo {pages} {2116} (\bibinfo {year} {2024})}\BibitemShut {NoStop}%
\bibitem [{\citenamefont {Aoyama}\ and\ \citenamefont {Ohgushi}(2024)}]{aoyama2024piezomagnetic}%
  \BibitemOpen
  \bibfield  {author} {\bibinfo {author} {\bibfnamefont {T.}~\bibnamefont {Aoyama}}\ and\ \bibinfo {author} {\bibfnamefont {K.}~\bibnamefont {Ohgushi}},\ }\href@noop {} {\bibfield  {journal} {\bibinfo  {journal} {Physical Review Materials}\ }\textbf {\bibinfo {volume} {8}},\ \bibinfo {pages} {L041402} (\bibinfo {year} {2024})}\BibitemShut {NoStop}%
\bibitem [{\citenamefont {Lee}\ \emph {et~al.}(2024)\citenamefont {Lee}, \citenamefont {Lee}, \citenamefont {Jung}, \citenamefont {Jung}, \citenamefont {Kim}, \citenamefont {Lee}, \citenamefont {Seok}, \citenamefont {Kim}, \citenamefont {Park}, \citenamefont {{\v{S}}mejkal} \emph {et~al.}}]{lee2024broken}%
  \BibitemOpen
  \bibfield  {author} {\bibinfo {author} {\bibfnamefont {S.}~\bibnamefont {Lee}}, \bibinfo {author} {\bibfnamefont {S.}~\bibnamefont {Lee}}, \bibinfo {author} {\bibfnamefont {S.}~\bibnamefont {Jung}}, \bibinfo {author} {\bibfnamefont {J.}~\bibnamefont {Jung}}, \bibinfo {author} {\bibfnamefont {D.}~\bibnamefont {Kim}}, \bibinfo {author} {\bibfnamefont {Y.}~\bibnamefont {Lee}}, \bibinfo {author} {\bibfnamefont {B.}~\bibnamefont {Seok}}, \bibinfo {author} {\bibfnamefont {J.}~\bibnamefont {Kim}}, \bibinfo {author} {\bibfnamefont {B.~G.}\ \bibnamefont {Park}}, \bibinfo {author} {\bibfnamefont {L.}~\bibnamefont {{\v{S}}mejkal}},  \emph {et~al.},\ }\href@noop {} {\bibfield  {journal} {\bibinfo  {journal} {Physical Review Letters}\ }\textbf {\bibinfo {volume} {132}},\ \bibinfo {pages} {036702} (\bibinfo {year} {2024})}\BibitemShut {NoStop}%
\bibitem [{\citenamefont {Feng}\ \emph {et~al.}(2022)\citenamefont {Feng}, \citenamefont {Zhou}, \citenamefont {{\v{S}}mejkal}, \citenamefont {Wu}, \citenamefont {Zhu}, \citenamefont {Guo}, \citenamefont {Gonz{\'a}lez-Hern{\'a}ndez}, \citenamefont {Wang}, \citenamefont {Yan}, \citenamefont {Qin} \emph {et~al.}}]{feng2022anomalous}%
  \BibitemOpen
  \bibfield  {author} {\bibinfo {author} {\bibfnamefont {Z.}~\bibnamefont {Feng}}, \bibinfo {author} {\bibfnamefont {X.}~\bibnamefont {Zhou}}, \bibinfo {author} {\bibfnamefont {L.}~\bibnamefont {{\v{S}}mejkal}}, \bibinfo {author} {\bibfnamefont {L.}~\bibnamefont {Wu}}, \bibinfo {author} {\bibfnamefont {Z.}~\bibnamefont {Zhu}}, \bibinfo {author} {\bibfnamefont {H.}~\bibnamefont {Guo}}, \bibinfo {author} {\bibfnamefont {R.}~\bibnamefont {Gonz{\'a}lez-Hern{\'a}ndez}}, \bibinfo {author} {\bibfnamefont {X.}~\bibnamefont {Wang}}, \bibinfo {author} {\bibfnamefont {H.}~\bibnamefont {Yan}}, \bibinfo {author} {\bibfnamefont {P.}~\bibnamefont {Qin}},  \emph {et~al.},\ }\href@noop {} {\bibfield  {journal} {\bibinfo  {journal} {Nature Electronics}\ }\textbf {\bibinfo {volume} {5}},\ \bibinfo {pages} {735} (\bibinfo {year} {2022})}\BibitemShut {NoStop}%
\bibitem [{\citenamefont {Leivisk{\"a}}\ \emph {et~al.}(2024)\citenamefont {Leivisk{\"a}}, \citenamefont {Rial}, \citenamefont {Bad'ura}, \citenamefont {Seeger}, \citenamefont {Kounta}, \citenamefont {Beckert}, \citenamefont {Kriegner}, \citenamefont {Joumard}, \citenamefont {Schmoranzerov{\'a}}, \citenamefont {Sinova} \emph {et~al.}}]{leiviska2024anisotropy}%
  \BibitemOpen
  \bibfield  {author} {\bibinfo {author} {\bibfnamefont {M.}~\bibnamefont {Leivisk{\"a}}}, \bibinfo {author} {\bibfnamefont {J.}~\bibnamefont {Rial}}, \bibinfo {author} {\bibfnamefont {A.}~\bibnamefont {Bad'ura}}, \bibinfo {author} {\bibfnamefont {R.~L.}\ \bibnamefont {Seeger}}, \bibinfo {author} {\bibfnamefont {I.}~\bibnamefont {Kounta}}, \bibinfo {author} {\bibfnamefont {S.}~\bibnamefont {Beckert}}, \bibinfo {author} {\bibfnamefont {D.}~\bibnamefont {Kriegner}}, \bibinfo {author} {\bibfnamefont {I.}~\bibnamefont {Joumard}}, \bibinfo {author} {\bibfnamefont {E.}~\bibnamefont {Schmoranzerov{\'a}}}, \bibinfo {author} {\bibfnamefont {J.}~\bibnamefont {Sinova}},  \emph {et~al.},\ }\href@noop {} {\bibfield  {journal} {\bibinfo  {journal} {Physical Review B}\ }\textbf {\bibinfo {volume} {109}},\ \bibinfo {pages} {224430} (\bibinfo {year} {2024})}\BibitemShut {NoStop}%
\bibitem [{\citenamefont {Sch{\"a}fer}\ \emph {et~al.}(1963)\citenamefont {Sch{\"a}fer}, \citenamefont {Schneidereit},\ and\ \citenamefont {Gerhardt}}]{schafer1963chemie}%
  \BibitemOpen
  \bibfield  {author} {\bibinfo {author} {\bibfnamefont {H.}~\bibnamefont {Sch{\"a}fer}}, \bibinfo {author} {\bibfnamefont {G.}~\bibnamefont {Schneidereit}}, \ and\ \bibinfo {author} {\bibfnamefont {W.}~\bibnamefont {Gerhardt}},\ }\href@noop {} {\bibfield  {journal} {\bibinfo  {journal} {Zeitschrift F{\"u}r Anorganische und Allgemeine Chemie}\ }\textbf {\bibinfo {volume} {319}},\ \bibinfo {pages} {327} (\bibinfo {year} {1963})}\BibitemShut {NoStop}%
\bibitem [{\citenamefont {Kameyama}\ \emph {et~al.}(1994)\citenamefont {Kameyama}, \citenamefont {Tsukada}, \citenamefont {Yahikozawa},\ and\ \citenamefont {Takasu}}]{kameyama1994surface}%
  \BibitemOpen
  \bibfield  {author} {\bibinfo {author} {\bibfnamefont {K.}~\bibnamefont {Kameyama}}, \bibinfo {author} {\bibfnamefont {K.}~\bibnamefont {Tsukada}}, \bibinfo {author} {\bibfnamefont {K.}~\bibnamefont {Yahikozawa}}, \ and\ \bibinfo {author} {\bibfnamefont {Y.}~\bibnamefont {Takasu}},\ }\href@noop {} {\bibfield  {journal} {\bibinfo  {journal} {Journal of the Electrochemical Society}\ }\textbf {\bibinfo {volume} {141}},\ \bibinfo {pages} {643} (\bibinfo {year} {1994})}\BibitemShut {NoStop}%
\bibitem [{\citenamefont {Bat'ko}\ \emph {et~al.}(1995)\citenamefont {Bat'ko}, \citenamefont {Flachbart}, \citenamefont {Somora},\ and\ \citenamefont {Vanick{\`y}}}]{bat1995design}%
  \BibitemOpen
  \bibfield  {author} {\bibinfo {author} {\bibfnamefont {I.}~\bibnamefont {Bat'ko}}, \bibinfo {author} {\bibfnamefont {K.}~\bibnamefont {Flachbart}}, \bibinfo {author} {\bibfnamefont {M.}~\bibnamefont {Somora}}, \ and\ \bibinfo {author} {\bibfnamefont {D.}~\bibnamefont {Vanick{\`y}}},\ }\href@noop {} {\bibfield  {journal} {\bibinfo  {journal} {Cryogenics}\ }\textbf {\bibinfo {volume} {35}},\ \bibinfo {pages} {105} (\bibinfo {year} {1995})}\BibitemShut {NoStop}%
\bibitem [{\citenamefont {Over}\ \emph {et~al.}(2000)\citenamefont {Over}, \citenamefont {Kim}, \citenamefont {Seitsonen}, \citenamefont {Wendt}, \citenamefont {Lundgren}, \citenamefont {Schmid}, \citenamefont {Varga}, \citenamefont {Morgante},\ and\ \citenamefont {Ertl}}]{over2000atomic}%
  \BibitemOpen
  \bibfield  {author} {\bibinfo {author} {\bibfnamefont {H.}~\bibnamefont {Over}}, \bibinfo {author} {\bibfnamefont {Y.~D.}\ \bibnamefont {Kim}}, \bibinfo {author} {\bibfnamefont {A.}~\bibnamefont {Seitsonen}}, \bibinfo {author} {\bibfnamefont {S.}~\bibnamefont {Wendt}}, \bibinfo {author} {\bibfnamefont {E.}~\bibnamefont {Lundgren}}, \bibinfo {author} {\bibfnamefont {M.}~\bibnamefont {Schmid}}, \bibinfo {author} {\bibfnamefont {P.}~\bibnamefont {Varga}}, \bibinfo {author} {\bibfnamefont {A.}~\bibnamefont {Morgante}}, \ and\ \bibinfo {author} {\bibfnamefont {G.}~\bibnamefont {Ertl}},\ }\href@noop {} {\bibfield  {journal} {\bibinfo  {journal} {Science}\ }\textbf {\bibinfo {volume} {287}},\ \bibinfo {pages} {1474} (\bibinfo {year} {2000})}\BibitemShut {NoStop}%
\bibitem [{\citenamefont {Berlijn}\ \emph {et~al.}(2017)\citenamefont {Berlijn}, \citenamefont {Snijders}, \citenamefont {Delaire}, \citenamefont {Zhou}, \citenamefont {Maier}, \citenamefont {Cao}, \citenamefont {Chi}, \citenamefont {Matsuda}, \citenamefont {Wang}, \citenamefont {Koehler} \emph {et~al.}}]{berlijn2017itinerant}%
  \BibitemOpen
  \bibfield  {author} {\bibinfo {author} {\bibfnamefont {T.}~\bibnamefont {Berlijn}}, \bibinfo {author} {\bibfnamefont {P.~C.}\ \bibnamefont {Snijders}}, \bibinfo {author} {\bibfnamefont {O.}~\bibnamefont {Delaire}}, \bibinfo {author} {\bibfnamefont {H.-D.}\ \bibnamefont {Zhou}}, \bibinfo {author} {\bibfnamefont {T.~A.}\ \bibnamefont {Maier}}, \bibinfo {author} {\bibfnamefont {H.-B.}\ \bibnamefont {Cao}}, \bibinfo {author} {\bibfnamefont {S.-X.}\ \bibnamefont {Chi}}, \bibinfo {author} {\bibfnamefont {M.}~\bibnamefont {Matsuda}}, \bibinfo {author} {\bibfnamefont {Y.}~\bibnamefont {Wang}}, \bibinfo {author} {\bibfnamefont {M.~R.}\ \bibnamefont {Koehler}},  \emph {et~al.},\ }\href@noop {} {\bibfield  {journal} {\bibinfo  {journal} {Physical Review Letters}\ }\textbf {\bibinfo {volume} {118}},\ \bibinfo {pages} {077201} (\bibinfo {year} {2017})}\BibitemShut {NoStop}%
\bibitem [{\citenamefont {Cordfunke}\ \emph {et~al.}(1989)\citenamefont {Cordfunke}, \citenamefont {Konings}, \citenamefont {Westrum~Jr},\ and\ \citenamefont {Shaviv}}]{cordfunke1989thermophysical}%
  \BibitemOpen
  \bibfield  {author} {\bibinfo {author} {\bibfnamefont {E.}~\bibnamefont {Cordfunke}}, \bibinfo {author} {\bibfnamefont {R.}~\bibnamefont {Konings}}, \bibinfo {author} {\bibfnamefont {E.~F.}\ \bibnamefont {Westrum~Jr}}, \ and\ \bibinfo {author} {\bibfnamefont {R.}~\bibnamefont {Shaviv}},\ }\href@noop {} {\bibfield  {journal} {\bibinfo  {journal} {Journal of Physics and Chemistry of Solids}\ }\textbf {\bibinfo {volume} {50}},\ \bibinfo {pages} {429} (\bibinfo {year} {1989})}\BibitemShut {NoStop}%
\bibitem [{\citenamefont {O'Neill}\ and\ \citenamefont {Nell}(1997)}]{o1997gibbs}%
  \BibitemOpen
  \bibfield  {author} {\bibinfo {author} {\bibfnamefont {H.~S.~C.}\ \bibnamefont {O'Neill}}\ and\ \bibinfo {author} {\bibfnamefont {H.}~\bibnamefont {Nell}, \bibfnamefont {Johan}},\ }\href@noop {} {\bibfield  {journal} {\bibinfo  {journal} {Geochimica et Cosmochimica Acta}\ }\textbf {\bibinfo {volume} {61}},\ \bibinfo {pages} {5279} (\bibinfo {year} {1997})}\BibitemShut {NoStop}%
\bibitem [{\citenamefont {Ryden}\ and\ \citenamefont {Lawson}(1970)}]{ryden1970magnetic}%
  \BibitemOpen
  \bibfield  {author} {\bibinfo {author} {\bibfnamefont {W.}~\bibnamefont {Ryden}}\ and\ \bibinfo {author} {\bibfnamefont {A.}~\bibnamefont {Lawson}},\ }\href@noop {} {\bibfield  {journal} {\bibinfo  {journal} {The Journal of Chemical Physics}\ }\textbf {\bibinfo {volume} {52}},\ \bibinfo {pages} {6058} (\bibinfo {year} {1970})}\BibitemShut {NoStop}%
\bibitem [{\citenamefont {Fletcher}\ \emph {et~al.}(1968)\citenamefont {Fletcher}, \citenamefont {Gardner}, \citenamefont {Greenfield}, \citenamefont {Holdoway},\ and\ \citenamefont {Rand}}]{fletcher1968magnetic}%
  \BibitemOpen
  \bibfield  {author} {\bibinfo {author} {\bibfnamefont {J.}~\bibnamefont {Fletcher}}, \bibinfo {author} {\bibfnamefont {W.}~\bibnamefont {Gardner}}, \bibinfo {author} {\bibfnamefont {B.}~\bibnamefont {Greenfield}}, \bibinfo {author} {\bibfnamefont {M.}~\bibnamefont {Holdoway}}, \ and\ \bibinfo {author} {\bibfnamefont {M.}~\bibnamefont {Rand}},\ }\href@noop {} {\bibfield  {journal} {\bibinfo  {journal} {Journal of the Chemical Society A: Inorganic, Physical, Theoretical}\ ,\ \bibinfo {pages} {653}} (\bibinfo {year} {1968})}\BibitemShut {NoStop}%
\bibitem [{\citenamefont {Lin}\ \emph {et~al.}(2004)\citenamefont {Lin}, \citenamefont {Huang}, \citenamefont {Lin}, \citenamefont {Lee}, \citenamefont {Liu}, \citenamefont {Zhang}, \citenamefont {Chen},\ and\ \citenamefont {Huang}}]{lin2004low}%
  \BibitemOpen
  \bibfield  {author} {\bibinfo {author} {\bibfnamefont {J.-J.}\ \bibnamefont {Lin}}, \bibinfo {author} {\bibfnamefont {S.}~\bibnamefont {Huang}}, \bibinfo {author} {\bibfnamefont {Y.}~\bibnamefont {Lin}}, \bibinfo {author} {\bibfnamefont {T.}~\bibnamefont {Lee}}, \bibinfo {author} {\bibfnamefont {H.}~\bibnamefont {Liu}}, \bibinfo {author} {\bibfnamefont {X.}~\bibnamefont {Zhang}}, \bibinfo {author} {\bibfnamefont {R.}~\bibnamefont {Chen}}, \ and\ \bibinfo {author} {\bibfnamefont {Y.}~\bibnamefont {Huang}},\ }\href@noop {} {\bibfield  {journal} {\bibinfo  {journal} {Journal of Physics: Condensed Matter}\ }\textbf {\bibinfo {volume} {16}},\ \bibinfo {pages} {8035} (\bibinfo {year} {2004})}\BibitemShut {NoStop}%
\bibitem [{\citenamefont {Glassford}\ and\ \citenamefont {Chelikowsky}(1994)}]{PhysRevB.49.7107}%
  \BibitemOpen
  \bibfield  {author} {\bibinfo {author} {\bibfnamefont {K.~M.}\ \bibnamefont {Glassford}}\ and\ \bibinfo {author} {\bibfnamefont {J.~R.}\ \bibnamefont {Chelikowsky}},\ }\href {\doibase 10.1103/PhysRevB.49.7107} {\bibfield  {journal} {\bibinfo  {journal} {Physical Review B}\ }\textbf {\bibinfo {volume} {49}},\ \bibinfo {pages} {7107} (\bibinfo {year} {1994})}\BibitemShut {NoStop}%
\bibitem [{\citenamefont {Touloukian}\ \emph {et~al.}(1977)\citenamefont {Touloukian}, \citenamefont {Kirby}, \citenamefont {Taylor},\ and\ \citenamefont {Lee}}]{touloukian1977thermophysical}%
  \BibitemOpen
  \bibfield  {author} {\bibinfo {author} {\bibfnamefont {Y.}~\bibnamefont {Touloukian}}, \bibinfo {author} {\bibfnamefont {R.}~\bibnamefont {Kirby}}, \bibinfo {author} {\bibfnamefont {E.}~\bibnamefont {Taylor}}, \ and\ \bibinfo {author} {\bibfnamefont {T.~Y.}\ \bibnamefont {Lee}},\ }\href@noop {} {\emph {\bibinfo {title} {Thermophysical properties of matter-the TPRC data series. Volume 13. Thermal expansion-nonmetallic solids.(Reannouncement). Data book}}},\ \bibinfo {type} {Tech. Rep.}\ (\bibinfo  {institution} {Purdue Univ., Lafayette, IN (United States). Thermophysical and Electronic~…},\ \bibinfo {year} {1977})\BibitemShut {NoStop}%
\bibitem [{\citenamefont {Hiraishi}\ \emph {et~al.}(2024)\citenamefont {Hiraishi}, \citenamefont {Okabe}, \citenamefont {Koda}, \citenamefont {Kadono}, \citenamefont {Muroi}, \citenamefont {Hirai},\ and\ \citenamefont {Hiroi}}]{hiraishi2024nonmagnetic}%
  \BibitemOpen
  \bibfield  {author} {\bibinfo {author} {\bibfnamefont {M.}~\bibnamefont {Hiraishi}}, \bibinfo {author} {\bibfnamefont {H.}~\bibnamefont {Okabe}}, \bibinfo {author} {\bibfnamefont {A.}~\bibnamefont {Koda}}, \bibinfo {author} {\bibfnamefont {R.}~\bibnamefont {Kadono}}, \bibinfo {author} {\bibfnamefont {T.}~\bibnamefont {Muroi}}, \bibinfo {author} {\bibfnamefont {D.}~\bibnamefont {Hirai}}, \ and\ \bibinfo {author} {\bibfnamefont {Z.}~\bibnamefont {Hiroi}},\ }\href@noop {} {\bibfield  {journal} {\bibinfo  {journal} {Physical Review Letters}\ }\textbf {\bibinfo {volume} {132}},\ \bibinfo {pages} {166702} (\bibinfo {year} {2024})}\BibitemShut {NoStop}%
\bibitem [{\citenamefont {Ke{\ss}ler}\ \emph {et~al.}(2024)\citenamefont {Ke{\ss}ler}, \citenamefont {Garcia-Gassull}, \citenamefont {Suter}, \citenamefont {Prokscha}, \citenamefont {Salman}, \citenamefont {Khalyavin}, \citenamefont {Manuel}, \citenamefont {Orlandi}, \citenamefont {Mazin}, \citenamefont {Valent{\'\i}} \emph {et~al.}}]{kessler2024absence}%
  \BibitemOpen
  \bibfield  {author} {\bibinfo {author} {\bibfnamefont {P.}~\bibnamefont {Ke{\ss}ler}}, \bibinfo {author} {\bibfnamefont {L.}~\bibnamefont {Garcia-Gassull}}, \bibinfo {author} {\bibfnamefont {A.}~\bibnamefont {Suter}}, \bibinfo {author} {\bibfnamefont {T.}~\bibnamefont {Prokscha}}, \bibinfo {author} {\bibfnamefont {Z.}~\bibnamefont {Salman}}, \bibinfo {author} {\bibfnamefont {D.}~\bibnamefont {Khalyavin}}, \bibinfo {author} {\bibfnamefont {P.}~\bibnamefont {Manuel}}, \bibinfo {author} {\bibfnamefont {F.}~\bibnamefont {Orlandi}}, \bibinfo {author} {\bibfnamefont {I.~I.}\ \bibnamefont {Mazin}}, \bibinfo {author} {\bibfnamefont {R.}~\bibnamefont {Valent{\'\i}}},  \emph {et~al.},\ }\href@noop {} {\bibfield  {journal} {\bibinfo  {journal} {npj Spintronics}\ }\textbf {\bibinfo {volume} {2}},\ \bibinfo {pages} {50} (\bibinfo {year} {2024})}\BibitemShut {NoStop}%
\bibitem [{\citenamefont {Pawula}\ \emph {et~al.}(2024)\citenamefont {Pawula}, \citenamefont {Fakih}, \citenamefont {Daou}, \citenamefont {H{\'e}bert}, \citenamefont {Mordvinova}, \citenamefont {Lebedev}, \citenamefont {Pelloquin},\ and\ \citenamefont {Maignan}}]{pawula2024multiband}%
  \BibitemOpen
  \bibfield  {author} {\bibinfo {author} {\bibfnamefont {F.}~\bibnamefont {Pawula}}, \bibinfo {author} {\bibfnamefont {A.}~\bibnamefont {Fakih}}, \bibinfo {author} {\bibfnamefont {R.}~\bibnamefont {Daou}}, \bibinfo {author} {\bibfnamefont {S.}~\bibnamefont {H{\'e}bert}}, \bibinfo {author} {\bibfnamefont {N.}~\bibnamefont {Mordvinova}}, \bibinfo {author} {\bibfnamefont {O.}~\bibnamefont {Lebedev}}, \bibinfo {author} {\bibfnamefont {D.}~\bibnamefont {Pelloquin}}, \ and\ \bibinfo {author} {\bibfnamefont {A.}~\bibnamefont {Maignan}},\ }\href@noop {} {\bibfield  {journal} {\bibinfo  {journal} {Physical Review B}\ }\textbf {\bibinfo {volume} {110}},\ \bibinfo {pages} {064432} (\bibinfo {year} {2024})}\BibitemShut {NoStop}%
\bibitem [{\citenamefont {Liu}\ \emph {et~al.}(2024)\citenamefont {Liu}, \citenamefont {Zhan}, \citenamefont {Li}, \citenamefont {Liu}, \citenamefont {Cheng}, \citenamefont {Shi}, \citenamefont {Deng}, \citenamefont {Zhang}, \citenamefont {Li}, \citenamefont {Ding} \emph {et~al.}}]{liu2024absence}%
  \BibitemOpen
  \bibfield  {author} {\bibinfo {author} {\bibfnamefont {J.}~\bibnamefont {Liu}}, \bibinfo {author} {\bibfnamefont {J.}~\bibnamefont {Zhan}}, \bibinfo {author} {\bibfnamefont {T.}~\bibnamefont {Li}}, \bibinfo {author} {\bibfnamefont {J.}~\bibnamefont {Liu}}, \bibinfo {author} {\bibfnamefont {S.}~\bibnamefont {Cheng}}, \bibinfo {author} {\bibfnamefont {Y.}~\bibnamefont {Shi}}, \bibinfo {author} {\bibfnamefont {L.}~\bibnamefont {Deng}}, \bibinfo {author} {\bibfnamefont {M.}~\bibnamefont {Zhang}}, \bibinfo {author} {\bibfnamefont {C.}~\bibnamefont {Li}}, \bibinfo {author} {\bibfnamefont {J.}~\bibnamefont {Ding}},  \emph {et~al.},\ }\href@noop {} {\bibfield  {journal} {\bibinfo  {journal} {Physical Review Letters}\ }\textbf {\bibinfo {volume} {133}},\ \bibinfo {pages} {176401} (\bibinfo {year} {2024})}\BibitemShut {NoStop}%
\bibitem [{\citenamefont {Wenzel}\ \emph {et~al.}(2025)\citenamefont {Wenzel}, \citenamefont {Uykur}, \citenamefont {R{\"o}{\ss}ler}, \citenamefont {Schmidt}, \citenamefont {Janson}, \citenamefont {Tiwari}, \citenamefont {Dressel},\ and\ \citenamefont {Tsirlin}}]{wenzel2025fermi}%
  \BibitemOpen
  \bibfield  {author} {\bibinfo {author} {\bibfnamefont {M.}~\bibnamefont {Wenzel}}, \bibinfo {author} {\bibfnamefont {E.}~\bibnamefont {Uykur}}, \bibinfo {author} {\bibfnamefont {S.}~\bibnamefont {R{\"o}{\ss}ler}}, \bibinfo {author} {\bibfnamefont {M.}~\bibnamefont {Schmidt}}, \bibinfo {author} {\bibfnamefont {O.}~\bibnamefont {Janson}}, \bibinfo {author} {\bibfnamefont {A.}~\bibnamefont {Tiwari}}, \bibinfo {author} {\bibfnamefont {M.}~\bibnamefont {Dressel}}, \ and\ \bibinfo {author} {\bibfnamefont {A.~A.}\ \bibnamefont {Tsirlin}},\ }\href@noop {} {\bibfield  {journal} {\bibinfo  {journal} {Physical Review B}\ }\textbf {\bibinfo {volume} {111}},\ \bibinfo {pages} {L041115} (\bibinfo {year} {2025})}\BibitemShut {NoStop}%
\bibitem [{\citenamefont {Zhu}\ \emph {et~al.}(2019)\citenamefont {Zhu}, \citenamefont {Strempfer}, \citenamefont {Rao}, \citenamefont {Occhialini}, \citenamefont {Pelliciari}, \citenamefont {Choi}, \citenamefont {Kawaguchi}, \citenamefont {You}, \citenamefont {Mitchell}, \citenamefont {Shao-Horn} \emph {et~al.}}]{zhu2019anomalous}%
  \BibitemOpen
  \bibfield  {author} {\bibinfo {author} {\bibfnamefont {Z.}~\bibnamefont {Zhu}}, \bibinfo {author} {\bibfnamefont {J.}~\bibnamefont {Strempfer}}, \bibinfo {author} {\bibfnamefont {R.}~\bibnamefont {Rao}}, \bibinfo {author} {\bibfnamefont {C.}~\bibnamefont {Occhialini}}, \bibinfo {author} {\bibfnamefont {J.}~\bibnamefont {Pelliciari}}, \bibinfo {author} {\bibfnamefont {Y.}~\bibnamefont {Choi}}, \bibinfo {author} {\bibfnamefont {T.}~\bibnamefont {Kawaguchi}}, \bibinfo {author} {\bibfnamefont {H.}~\bibnamefont {You}}, \bibinfo {author} {\bibfnamefont {J.}~\bibnamefont {Mitchell}}, \bibinfo {author} {\bibfnamefont {Y.}~\bibnamefont {Shao-Horn}},  \emph {et~al.},\ }\href@noop {} {\bibfield  {journal} {\bibinfo  {journal} {Physical Review Letters}\ }\textbf {\bibinfo {volume} {122}},\ \bibinfo {pages} {017202} (\bibinfo {year} {2019})}\BibitemShut {NoStop}%
\bibitem [{\citenamefont {Tschirner}\ \emph {et~al.}(2023)\citenamefont {Tschirner}, \citenamefont {Ke{\ss}ler}, \citenamefont {Gonzalez~Betancourt}, \citenamefont {Kotte}, \citenamefont {Kriegner}, \citenamefont {B{\"u}chner}, \citenamefont {Dufouleur}, \citenamefont {Kamp}, \citenamefont {Jovic}, \citenamefont {Smejkal} \emph {et~al.}}]{tschirner2023saturation}%
  \BibitemOpen
  \bibfield  {author} {\bibinfo {author} {\bibfnamefont {T.}~\bibnamefont {Tschirner}}, \bibinfo {author} {\bibfnamefont {P.}~\bibnamefont {Ke{\ss}ler}}, \bibinfo {author} {\bibfnamefont {R.~D.}\ \bibnamefont {Gonzalez~Betancourt}}, \bibinfo {author} {\bibfnamefont {T.}~\bibnamefont {Kotte}}, \bibinfo {author} {\bibfnamefont {D.}~\bibnamefont {Kriegner}}, \bibinfo {author} {\bibfnamefont {B.}~\bibnamefont {B{\"u}chner}}, \bibinfo {author} {\bibfnamefont {J.}~\bibnamefont {Dufouleur}}, \bibinfo {author} {\bibfnamefont {M.}~\bibnamefont {Kamp}}, \bibinfo {author} {\bibfnamefont {V.}~\bibnamefont {Jovic}}, \bibinfo {author} {\bibfnamefont {L.}~\bibnamefont {Smejkal}},  \emph {et~al.},\ }\href@noop {} {\bibfield  {journal} {\bibinfo  {journal} {APL Materials}\ }\textbf {\bibinfo {volume} {11}} (\bibinfo {year} {2023})}\BibitemShut {NoStop}%
\bibitem [{\citenamefont {Jeong}\ \emph {et~al.}(2025{\natexlab{a}})\citenamefont {Jeong}, \citenamefont {Lee}, \citenamefont {Lin}, \citenamefont {Yang}, \citenamefont {Choi}, \citenamefont {Oh}, \citenamefont {Song}, \citenamefont {Nair}, \citenamefont {Choudhary}, \citenamefont {Parikh} \emph {et~al.}}]{jeong2025metallicity}%
  \BibitemOpen
  \bibfield  {author} {\bibinfo {author} {\bibfnamefont {S.~G.}\ \bibnamefont {Jeong}}, \bibinfo {author} {\bibfnamefont {S.}~\bibnamefont {Lee}}, \bibinfo {author} {\bibfnamefont {B.}~\bibnamefont {Lin}}, \bibinfo {author} {\bibfnamefont {Z.}~\bibnamefont {Yang}}, \bibinfo {author} {\bibfnamefont {I.~H.}\ \bibnamefont {Choi}}, \bibinfo {author} {\bibfnamefont {J.~Y.}\ \bibnamefont {Oh}}, \bibinfo {author} {\bibfnamefont {S.}~\bibnamefont {Song}}, \bibinfo {author} {\bibfnamefont {S.}~\bibnamefont {Nair}}, \bibinfo {author} {\bibfnamefont {R.}~\bibnamefont {Choudhary}}, \bibinfo {author} {\bibfnamefont {J.}~\bibnamefont {Parikh}},  \emph {et~al.},\ }\href@noop {} {\bibfield  {journal} {\bibinfo  {journal} {arXiv preprint arXiv:2501.11204}\ } (\bibinfo {year} {2025}{\natexlab{a}})}\BibitemShut {NoStop}%
\bibitem [{\citenamefont {Karube}\ \emph {et~al.}(2022)\citenamefont {Karube}, \citenamefont {Tanaka}, \citenamefont {Sugawara}, \citenamefont {Kadoguchi}, \citenamefont {Kohda},\ and\ \citenamefont {Nitta}}]{karube2022observation}%
  \BibitemOpen
  \bibfield  {author} {\bibinfo {author} {\bibfnamefont {S.}~\bibnamefont {Karube}}, \bibinfo {author} {\bibfnamefont {T.}~\bibnamefont {Tanaka}}, \bibinfo {author} {\bibfnamefont {D.}~\bibnamefont {Sugawara}}, \bibinfo {author} {\bibfnamefont {N.}~\bibnamefont {Kadoguchi}}, \bibinfo {author} {\bibfnamefont {M.}~\bibnamefont {Kohda}}, \ and\ \bibinfo {author} {\bibfnamefont {J.}~\bibnamefont {Nitta}},\ }\href@noop {} {\bibfield  {journal} {\bibinfo  {journal} {Physical Review Letters}\ }\textbf {\bibinfo {volume} {129}},\ \bibinfo {pages} {137201} (\bibinfo {year} {2022})}\BibitemShut {NoStop}%
\bibitem [{\citenamefont {Bai}\ \emph {et~al.}(2023)\citenamefont {Bai}, \citenamefont {Zhang}, \citenamefont {Zhou}, \citenamefont {Chen}, \citenamefont {Wan}, \citenamefont {Han}, \citenamefont {Zhu}, \citenamefont {Liang}, \citenamefont {Su}, \citenamefont {Han} \emph {et~al.}}]{bai2023efficient}%
  \BibitemOpen
  \bibfield  {author} {\bibinfo {author} {\bibfnamefont {H.}~\bibnamefont {Bai}}, \bibinfo {author} {\bibfnamefont {Y.}~\bibnamefont {Zhang}}, \bibinfo {author} {\bibfnamefont {Y.}~\bibnamefont {Zhou}}, \bibinfo {author} {\bibfnamefont {P.}~\bibnamefont {Chen}}, \bibinfo {author} {\bibfnamefont {C.}~\bibnamefont {Wan}}, \bibinfo {author} {\bibfnamefont {L.}~\bibnamefont {Han}}, \bibinfo {author} {\bibfnamefont {W.}~\bibnamefont {Zhu}}, \bibinfo {author} {\bibfnamefont {S.}~\bibnamefont {Liang}}, \bibinfo {author} {\bibfnamefont {Y.}~\bibnamefont {Su}}, \bibinfo {author} {\bibfnamefont {X.}~\bibnamefont {Han}},  \emph {et~al.},\ }\href@noop {} {\bibfield  {journal} {\bibinfo  {journal} {Physical Review Letters}\ }\textbf {\bibinfo {volume} {130}},\ \bibinfo {pages} {216701} (\bibinfo {year} {2023})}\BibitemShut {NoStop}%
\bibitem [{\citenamefont {Feng}\ \emph {et~al.}(2024)\citenamefont {Feng}, \citenamefont {Bai}, \citenamefont {Fan}, \citenamefont {Guo}, \citenamefont {Zhang}, \citenamefont {Chai}, \citenamefont {Wang}, \citenamefont {Xue}, \citenamefont {Song},\ and\ \citenamefont {Fan}}]{feng2024incommensurate}%
  \BibitemOpen
  \bibfield  {author} {\bibinfo {author} {\bibfnamefont {X.}~\bibnamefont {Feng}}, \bibinfo {author} {\bibfnamefont {H.}~\bibnamefont {Bai}}, \bibinfo {author} {\bibfnamefont {X.}~\bibnamefont {Fan}}, \bibinfo {author} {\bibfnamefont {M.}~\bibnamefont {Guo}}, \bibinfo {author} {\bibfnamefont {Z.}~\bibnamefont {Zhang}}, \bibinfo {author} {\bibfnamefont {G.}~\bibnamefont {Chai}}, \bibinfo {author} {\bibfnamefont {T.}~\bibnamefont {Wang}}, \bibinfo {author} {\bibfnamefont {D.}~\bibnamefont {Xue}}, \bibinfo {author} {\bibfnamefont {C.}~\bibnamefont {Song}}, \ and\ \bibinfo {author} {\bibfnamefont {X.}~\bibnamefont {Fan}},\ }\href@noop {} {\bibfield  {journal} {\bibinfo  {journal} {Physical Review Letters}\ }\textbf {\bibinfo {volume} {132}},\ \bibinfo {pages} {086701} (\bibinfo {year} {2024})}\BibitemShut {NoStop}%
\bibitem [{\citenamefont {Weber}\ \emph {et~al.}(2024)\citenamefont {Weber}, \citenamefont {Wust}, \citenamefont {Haag}, \citenamefont {Akashdeep}, \citenamefont {Leckron}, \citenamefont {Schmitt}, \citenamefont {Ramos}, \citenamefont {Kikkawa}, \citenamefont {Saitoh}, \citenamefont {Kl{\"a}ui} \emph {et~al.}}]{weber2024all}%
  \BibitemOpen
  \bibfield  {author} {\bibinfo {author} {\bibfnamefont {M.}~\bibnamefont {Weber}}, \bibinfo {author} {\bibfnamefont {S.}~\bibnamefont {Wust}}, \bibinfo {author} {\bibfnamefont {L.}~\bibnamefont {Haag}}, \bibinfo {author} {\bibfnamefont {A.}~\bibnamefont {Akashdeep}}, \bibinfo {author} {\bibfnamefont {K.}~\bibnamefont {Leckron}}, \bibinfo {author} {\bibfnamefont {C.}~\bibnamefont {Schmitt}}, \bibinfo {author} {\bibfnamefont {R.}~\bibnamefont {Ramos}}, \bibinfo {author} {\bibfnamefont {T.}~\bibnamefont {Kikkawa}}, \bibinfo {author} {\bibfnamefont {E.}~\bibnamefont {Saitoh}}, \bibinfo {author} {\bibfnamefont {M.}~\bibnamefont {Kl{\"a}ui}},  \emph {et~al.},\ }\href@noop {} {\bibfield  {journal} {\bibinfo  {journal} {arXiv preprint arXiv:2408.05187}\ } (\bibinfo {year} {2024})}\BibitemShut {NoStop}%
\bibitem [{\citenamefont {Jeong}\ \emph {et~al.}(2025{\natexlab{b}})\citenamefont {Jeong}, \citenamefont {Choi}, \citenamefont {Nair}, \citenamefont {Buiarelli}, \citenamefont {Pourbahari}, \citenamefont {Oh}, \citenamefont {Bassim}, \citenamefont {Hirai}, \citenamefont {Seo}, \citenamefont {Choi}, \citenamefont {Fernandes}, \citenamefont {Birol}, \citenamefont {Zhao}, \citenamefont {Lee},\ and\ \citenamefont {Jalan}}]{jeong2025altermagnetic}%
  \BibitemOpen
  \bibfield  {author} {\bibinfo {author} {\bibfnamefont {S.~G.}\ \bibnamefont {Jeong}}, \bibinfo {author} {\bibfnamefont {I.~H.}\ \bibnamefont {Choi}}, \bibinfo {author} {\bibfnamefont {S.}~\bibnamefont {Nair}}, \bibinfo {author} {\bibfnamefont {L.}~\bibnamefont {Buiarelli}}, \bibinfo {author} {\bibfnamefont {B.}~\bibnamefont {Pourbahari}}, \bibinfo {author} {\bibfnamefont {J.~Y.}\ \bibnamefont {Oh}}, \bibinfo {author} {\bibfnamefont {N.}~\bibnamefont {Bassim}}, \bibinfo {author} {\bibfnamefont {D.}~\bibnamefont {Hirai}}, \bibinfo {author} {\bibfnamefont {A.}~\bibnamefont {Seo}}, \bibinfo {author} {\bibfnamefont {W.~S.}\ \bibnamefont {Choi}}, \bibinfo {author} {\bibfnamefont {R.~M.}\ \bibnamefont {Fernandes}}, \bibinfo {author} {\bibfnamefont {T.}~\bibnamefont {Birol}}, \bibinfo {author} {\bibfnamefont {L.}~\bibnamefont {Zhao}}, \bibinfo {author} {\bibfnamefont {J.~S.}\ \bibnamefont {Lee}}, \ and\ \bibinfo {author} {\bibfnamefont {B.}~\bibnamefont {Jalan}},\ }\href {https://arxiv.org/abs/2405.05838} {\enquote
  {\bibinfo {title} {Altermagnetic polar metallic phase in ultra-thin epitaxially-strained ruo2 films},}\ } (\bibinfo {year} {2025}{\natexlab{b}}),\ \Eprint {http://arxiv.org/abs/2405.05838} {arXiv:2405.05838 [cond-mat.mtrl-sci]} \BibitemShut {NoStop}%
\bibitem [{\citenamefont {P{\'e}rez-Mato}\ \emph {et~al.}(2015)\citenamefont {P{\'e}rez-Mato}, \citenamefont {Gallego}, \citenamefont {Tasci}, \citenamefont {Elcoro}, \citenamefont {de~la Flor},\ and\ \citenamefont {Aroyo}}]{PerezMato2015}%
  \BibitemOpen
  \bibfield  {author} {\bibinfo {author} {\bibfnamefont {J.}~\bibnamefont {P{\'e}rez-Mato}}, \bibinfo {author} {\bibfnamefont {S.}~\bibnamefont {Gallego}}, \bibinfo {author} {\bibfnamefont {E.}~\bibnamefont {Tasci}}, \bibinfo {author} {\bibfnamefont {L.}~\bibnamefont {Elcoro}}, \bibinfo {author} {\bibfnamefont {G.}~\bibnamefont {de~la Flor}}, \ and\ \bibinfo {author} {\bibfnamefont {M.}~\bibnamefont {Aroyo}},\ }\href@noop {} {\bibfield  {journal} {\bibinfo  {journal} {Annual Review of Materials Research}\ }\textbf {\bibinfo {volume} {45}},\ \bibinfo {pages} {217} (\bibinfo {year} {2015})}\BibitemShut {NoStop}%
\bibitem [{\citenamefont {Bazhan}\ and\ \citenamefont {Bazan}(1975)}]{Bazan1975weak}%
  \BibitemOpen
  \bibfield  {author} {\bibinfo {author} {\bibfnamefont {A.}~\bibnamefont {Bazhan}}\ and\ \bibinfo {author} {\bibfnamefont {C.}~\bibnamefont {Bazan}},\ }\href@noop {} {\bibfield  {journal} {\bibinfo  {journal} {Journal of Experimental and Theoretical Physics}\ }\textbf {\bibinfo {volume} {42}} (\bibinfo {year} {1975})}\BibitemShut {NoStop}%
\bibitem [{SM()}]{SM}%
  \BibitemOpen
  \href@noop {} {\bibinfo  {journal} {See Supplemental Material at [URL will be inserted by publisher] for additional experimental details and measurements on multiple samples}\ }\BibitemShut {NoStop}%
\bibitem [{\citenamefont {Perfetti}(2017)}]{perfetti2017cantilever}%
  \BibitemOpen
\bibfield  {journal} {  }\bibfield  {author} {\bibinfo {author} {\bibfnamefont {M.}~\bibnamefont {Perfetti}},\ }\href@noop {} {\bibfield  {journal} {\bibinfo  {journal} {Coordination Chemistry Reviews}\ }\textbf {\bibinfo {volume} {348}},\ \bibinfo {pages} {171} (\bibinfo {year} {2017})}\BibitemShut {NoStop}%
\bibitem [{\citenamefont {Komatsubara}\ \emph {et~al.}(1963)\citenamefont {Komatsubara}, \citenamefont {Murakami},\ and\ \citenamefont {Hirahara}}]{komatsubara1963magnetic}%
  \BibitemOpen
  \bibfield  {author} {\bibinfo {author} {\bibfnamefont {T.}~\bibnamefont {Komatsubara}}, \bibinfo {author} {\bibfnamefont {M.}~\bibnamefont {Murakami}}, \ and\ \bibinfo {author} {\bibfnamefont {E.}~\bibnamefont {Hirahara}},\ }\href@noop {} {\bibfield  {journal} {\bibinfo  {journal} {Journal of the Physical Society of Japan}\ }\textbf {\bibinfo {volume} {18}},\ \bibinfo {pages} {356} (\bibinfo {year} {1963})}\BibitemShut {NoStop}%
\bibitem [{\citenamefont {Blundell}(2001)}]{blundell2001magnetism}%
  \BibitemOpen
  \bibfield  {author} {\bibinfo {author} {\bibfnamefont {S.}~\bibnamefont {Blundell}},\ }\href@noop {} {\emph {\bibinfo {title} {Magnetism in condensed matter}}}\ (\bibinfo  {publisher} {OUP Oxford},\ \bibinfo {year} {2001})\BibitemShut {NoStop}%
\bibitem [{\citenamefont {Lidiard}(1953)}]{Lidiard1953}%
  \BibitemOpen
  \bibfield  {author} {\bibinfo {author} {\bibfnamefont {A.}~\bibnamefont {Lidiard}},\ }\href@noop {} {\bibfield  {journal} {\bibinfo  {journal} {Proceedings of the Royal Society A}\ }\textbf {\bibinfo {volume} {224}},\ \bibinfo {pages} {161} (\bibinfo {year} {1953})}\BibitemShut {NoStop}%
\bibitem [{\citenamefont {Kiefer}\ \emph {et~al.}(2025)\citenamefont {Kiefer}, \citenamefont {Wirth}, \citenamefont {Bertin}, \citenamefont {Becker}, \citenamefont {Bohatý}, \citenamefont {Schmalzl}, \citenamefont {Stunault}, \citenamefont {Rodríguez-Velamazan}, \citenamefont {Fabelo},\ and\ \citenamefont {Braden}}]{Kiefer2025}%
  \BibitemOpen
  \bibfield  {author} {\bibinfo {author} {\bibfnamefont {L.}~\bibnamefont {Kiefer}}, \bibinfo {author} {\bibfnamefont {F.}~\bibnamefont {Wirth}}, \bibinfo {author} {\bibfnamefont {A.}~\bibnamefont {Bertin}}, \bibinfo {author} {\bibfnamefont {P.}~\bibnamefont {Becker}}, \bibinfo {author} {\bibfnamefont {L.}~\bibnamefont {Bohatý}}, \bibinfo {author} {\bibfnamefont {K.}~\bibnamefont {Schmalzl}}, \bibinfo {author} {\bibfnamefont {A.}~\bibnamefont {Stunault}}, \bibinfo {author} {\bibfnamefont {J.~A.}\ \bibnamefont {Rodríguez-Velamazan}}, \bibinfo {author} {\bibfnamefont {O.}~\bibnamefont {Fabelo}}, \ and\ \bibinfo {author} {\bibfnamefont {M.}~\bibnamefont {Braden}},\ }\href@noop {} {\bibfield  {journal} {\bibinfo  {journal} {Journal of Physics: Condensed Matter}\ }\textbf {\bibinfo {volume} {37}},\ \bibinfo {pages} {135801} (\bibinfo {year} {2025})}\BibitemShut {NoStop}%
\bibitem [{\citenamefont {Graebner}\ \emph {et~al.}(1976)\citenamefont {Graebner}, \citenamefont {Greiner},\ and\ \citenamefont {Ryden}}]{graebner1976magnetothermal}%
  \BibitemOpen
  \bibfield  {author} {\bibinfo {author} {\bibfnamefont {J.}~\bibnamefont {Graebner}}, \bibinfo {author} {\bibfnamefont {E.}~\bibnamefont {Greiner}}, \ and\ \bibinfo {author} {\bibfnamefont {W.}~\bibnamefont {Ryden}},\ }\href@noop {} {\bibfield  {journal} {\bibinfo  {journal} {Physical Review B}\ }\textbf {\bibinfo {volume} {13}},\ \bibinfo {pages} {2426} (\bibinfo {year} {1976})}\BibitemShut {NoStop}%
\bibitem [{\citenamefont {Shoenberg}(1984)}]{Shoenberg}%
  \BibitemOpen
  \bibfield  {author} {\bibinfo {author} {\bibfnamefont {D.}~\bibnamefont {Shoenberg}},\ }\href@noop {} {\emph {\bibinfo {title} {Magnetic Oscillations in Metals}}}\ (\bibinfo  {publisher} {Cambridge University Press},\ \bibinfo {year} {1984})\BibitemShut {NoStop}%
\bibitem [{\citenamefont {Ahn}\ \emph {et~al.}(2019)\citenamefont {Ahn}, \citenamefont {Hariki}, \citenamefont {Lee},\ and\ \citenamefont {Kune{\v{s}}}}]{ahn2019antiferromagnetism}%
  \BibitemOpen
  \bibfield  {author} {\bibinfo {author} {\bibfnamefont {K.-H.}\ \bibnamefont {Ahn}}, \bibinfo {author} {\bibfnamefont {A.}~\bibnamefont {Hariki}}, \bibinfo {author} {\bibfnamefont {K.-W.}\ \bibnamefont {Lee}}, \ and\ \bibinfo {author} {\bibfnamefont {J.}~\bibnamefont {Kune{\v{s}}}},\ }\href@noop {} {\bibfield  {journal} {\bibinfo  {journal} {Physical Review B}\ }\textbf {\bibinfo {volume} {99}},\ \bibinfo {pages} {184432} (\bibinfo {year} {2019})}\BibitemShut {NoStop}%
\bibitem [{\citenamefont {Huang}\ \emph {et~al.}(2024)\citenamefont {Huang}, \citenamefont {Lai}, \citenamefont {Zhan}, \citenamefont {Yu}, \citenamefont {Chen}, \citenamefont {Liu}, \citenamefont {Chen},\ and\ \citenamefont {Sun}}]{huang2024ab}%
  \BibitemOpen
  \bibfield  {author} {\bibinfo {author} {\bibfnamefont {Y.}~\bibnamefont {Huang}}, \bibinfo {author} {\bibfnamefont {J.}~\bibnamefont {Lai}}, \bibinfo {author} {\bibfnamefont {J.}~\bibnamefont {Zhan}}, \bibinfo {author} {\bibfnamefont {T.}~\bibnamefont {Yu}}, \bibinfo {author} {\bibfnamefont {R.}~\bibnamefont {Chen}}, \bibinfo {author} {\bibfnamefont {P.}~\bibnamefont {Liu}}, \bibinfo {author} {\bibfnamefont {X.-Q.}\ \bibnamefont {Chen}}, \ and\ \bibinfo {author} {\bibfnamefont {Y.}~\bibnamefont {Sun}},\ }\href@noop {} {\bibfield  {journal} {\bibinfo  {journal} {Physical Review B}\ }\textbf {\bibinfo {volume} {110}},\ \bibinfo {pages} {144410} (\bibinfo {year} {2024})}\BibitemShut {NoStop}%
\bibitem [{\citenamefont {Marcus}\ and\ \citenamefont {Butler}(1968)}]{marcus1968measurement}%
  \BibitemOpen
  \bibfield  {author} {\bibinfo {author} {\bibfnamefont {S.}~\bibnamefont {Marcus}}\ and\ \bibinfo {author} {\bibfnamefont {S.}~\bibnamefont {Butler}},\ }\href@noop {} {\bibfield  {journal} {\bibinfo  {journal} {Physics Letters A}\ }\textbf {\bibinfo {volume} {26}},\ \bibinfo {pages} {518} (\bibinfo {year} {1968})}\BibitemShut {NoStop}%
\bibitem [{\citenamefont {Samanta}\ \emph {et~al.}(2024)\citenamefont {Samanta}, \citenamefont {Jiang}, \citenamefont {Paudel}, \citenamefont {Shao},\ and\ \citenamefont {Tsymbal}}]{samanta2024tunneling}%
  \BibitemOpen
  \bibfield  {author} {\bibinfo {author} {\bibfnamefont {K.}~\bibnamefont {Samanta}}, \bibinfo {author} {\bibfnamefont {Y.-Y.}\ \bibnamefont {Jiang}}, \bibinfo {author} {\bibfnamefont {T.~R.}\ \bibnamefont {Paudel}}, \bibinfo {author} {\bibfnamefont {D.-F.}\ \bibnamefont {Shao}}, \ and\ \bibinfo {author} {\bibfnamefont {E.~Y.}\ \bibnamefont {Tsymbal}},\ }\href@noop {} {\bibfield  {journal} {\bibinfo  {journal} {Physical Review B}\ }\textbf {\bibinfo {volume} {109}},\ \bibinfo {pages} {174407} (\bibinfo {year} {2024})}\BibitemShut {NoStop}%
\bibitem [{\citenamefont {Fedchenko}\ \emph {et~al.}(2024)\citenamefont {Fedchenko}, \citenamefont {Min{\'a}r}, \citenamefont {Akashdeep}, \citenamefont {D’Souza}, \citenamefont {Vasilyev}, \citenamefont {Tkach}, \citenamefont {Odenbreit}, \citenamefont {Nguyen}, \citenamefont {Kutnyakhov}, \citenamefont {Wind} \emph {et~al.}}]{fedchenko2024observation}%
  \BibitemOpen
  \bibfield  {author} {\bibinfo {author} {\bibfnamefont {O.}~\bibnamefont {Fedchenko}}, \bibinfo {author} {\bibfnamefont {J.}~\bibnamefont {Min{\'a}r}}, \bibinfo {author} {\bibfnamefont {A.}~\bibnamefont {Akashdeep}}, \bibinfo {author} {\bibfnamefont {S.~W.}\ \bibnamefont {D’Souza}}, \bibinfo {author} {\bibfnamefont {D.}~\bibnamefont {Vasilyev}}, \bibinfo {author} {\bibfnamefont {O.}~\bibnamefont {Tkach}}, \bibinfo {author} {\bibfnamefont {L.}~\bibnamefont {Odenbreit}}, \bibinfo {author} {\bibfnamefont {Q.}~\bibnamefont {Nguyen}}, \bibinfo {author} {\bibfnamefont {D.}~\bibnamefont {Kutnyakhov}}, \bibinfo {author} {\bibfnamefont {N.}~\bibnamefont {Wind}},  \emph {et~al.},\ }\href@noop {} {\bibfield  {journal} {\bibinfo  {journal} {Science Advances}\ }\textbf {\bibinfo {volume} {10}},\ \bibinfo {pages} {eadj4883} (\bibinfo {year} {2024})}\BibitemShut {NoStop}%
\bibitem [{\citenamefont {{\v{S}}mejkal}\ and\ \citenamefont {Sinova}(2020)}]{smejkal2020}%
  \BibitemOpen
  \bibfield  {author} {\bibinfo {author} {\bibfnamefont {G.-H. R. J.~T.}\ \bibnamefont {{\v{S}}mejkal}, \bibfnamefont {Libor}}\ and\ \bibinfo {author} {\bibfnamefont {J.}~\bibnamefont {Sinova}},\ }\href@noop {} {\bibfield  {journal} {\bibinfo  {journal} {Science Advances}\ }\textbf {\bibinfo {volume} {6}},\ \bibinfo {pages} {eaaz8809} (\bibinfo {year} {2020})}\BibitemShut {NoStop}%
\bibitem [{\citenamefont {Glassford}\ and\ \citenamefont {Chelikowsky}(1993)}]{glassford1993electronic}%
  \BibitemOpen
  \bibfield  {author} {\bibinfo {author} {\bibfnamefont {K.~M.}\ \bibnamefont {Glassford}}\ and\ \bibinfo {author} {\bibfnamefont {J.~R.}\ \bibnamefont {Chelikowsky}},\ }\href@noop {} {\bibfield  {journal} {\bibinfo  {journal} {Physical Review B}\ }\textbf {\bibinfo {volume} {47}},\ \bibinfo {pages} {1732} (\bibinfo {year} {1993})}\BibitemShut {NoStop}%
\bibitem [{\citenamefont {Stoner}(1936)}]{Stoner1936}%
  \BibitemOpen
  \bibfield  {author} {\bibinfo {author} {\bibfnamefont {E.~C.}\ \bibnamefont {Stoner}},\ }\href@noop {} {\bibfield  {journal} {\bibinfo  {journal} {Proceedings of the Royal Society A}\ }\textbf {\bibinfo {volume} {154}},\ \bibinfo {pages} {656} (\bibinfo {year} {1936})}\BibitemShut {NoStop}%
\bibitem [{\citenamefont {Kriessman}\ and\ \citenamefont {Callen}(1954)}]{Kriessman1954}%
  \BibitemOpen
  \bibfield  {author} {\bibinfo {author} {\bibfnamefont {C.}~\bibnamefont {Kriessman}}\ and\ \bibinfo {author} {\bibfnamefont {H.}~\bibnamefont {Callen}},\ }\href@noop {} {\bibfield  {journal} {\bibinfo  {journal} {Physical Review}\ }\textbf {\bibinfo {volume} {94}},\ \bibinfo {pages} {837} (\bibinfo {year} {1954})}\BibitemShut {NoStop}%
\bibitem [{\citenamefont {Cox}\ \emph {et~al.}(1986)\citenamefont {Cox}, \citenamefont {Goodenough}, \citenamefont {Tavener}, \citenamefont {Telles},\ and\ \citenamefont {Egdell}}]{cox1986electronic}%
  \BibitemOpen
  \bibfield  {author} {\bibinfo {author} {\bibfnamefont {P.}~\bibnamefont {Cox}}, \bibinfo {author} {\bibfnamefont {J.}~\bibnamefont {Goodenough}}, \bibinfo {author} {\bibfnamefont {P.}~\bibnamefont {Tavener}}, \bibinfo {author} {\bibfnamefont {D.}~\bibnamefont {Telles}}, \ and\ \bibinfo {author} {\bibfnamefont {R.}~\bibnamefont {Egdell}},\ }\href@noop {} {\bibfield  {journal} {\bibinfo  {journal} {Journal of Solid State Chemistry}\ }\textbf {\bibinfo {volume} {62}},\ \bibinfo {pages} {360} (\bibinfo {year} {1986})}\BibitemShut {NoStop}%
\bibitem [{\citenamefont {Shimizu}\ and\ \citenamefont {Takahashi}(1960)}]{shimizu1960magnetic}%
  \BibitemOpen
  \bibfield  {author} {\bibinfo {author} {\bibfnamefont {M.}~\bibnamefont {Shimizu}}\ and\ \bibinfo {author} {\bibfnamefont {T.}~\bibnamefont {Takahashi}},\ }\href@noop {} {\bibfield  {journal} {\bibinfo  {journal} {Journal of the Physical Society of Japan}\ }\textbf {\bibinfo {volume} {15}},\ \bibinfo {pages} {2236} (\bibinfo {year} {1960})}\BibitemShut {NoStop}%
\bibitem [{\citenamefont {Mattheiss}(1976)}]{Mattheiss1976}%
  \BibitemOpen
  \bibfield  {author} {\bibinfo {author} {\bibfnamefont {L.~F.}\ \bibnamefont {Mattheiss}},\ }\href@noop {} {\bibfield  {journal} {\bibinfo  {journal} {Physical Review B}\ }\textbf {\bibinfo {volume} {13}},\ \bibinfo {pages} {2433} (\bibinfo {year} {1976})}\BibitemShut {NoStop}%
\bibitem [{\citenamefont {Galoshina}(1974)}]{Galoshina1974}%
  \BibitemOpen
  \bibfield  {author} {\bibinfo {author} {\bibfnamefont {E.~V.}\ \bibnamefont {Galoshina}},\ }\href@noop {} {\bibfield  {journal} {\bibinfo  {journal} {Soviet Physics Uspekhi}\ }\textbf {\bibinfo {volume} {17}},\ \bibinfo {pages} {345} (\bibinfo {year} {1974})}\BibitemShut {NoStop}%
\bibitem [{\citenamefont {Shimizu}(1981)}]{Shimizu1981}%
  \BibitemOpen
  \bibfield  {author} {\bibinfo {author} {\bibfnamefont {M.}~\bibnamefont {Shimizu}},\ }\href@noop {} {\bibfield  {journal} {\bibinfo  {journal} {Reports on Progress in Physics}\ }\textbf {\bibinfo {volume} {44}},\ \bibinfo {pages} {331} (\bibinfo {year} {1981})}\BibitemShut {NoStop}%
\bibitem [{\citenamefont {Smolyanyuk}\ \emph {et~al.}(2024)\citenamefont {Smolyanyuk}, \citenamefont {Mazin}, \citenamefont {Garcia-Gassull},\ and\ \citenamefont {Valent{\'\i}}}]{smolyanyuk2024fragility}%
  \BibitemOpen
  \bibfield  {author} {\bibinfo {author} {\bibfnamefont {A.}~\bibnamefont {Smolyanyuk}}, \bibinfo {author} {\bibfnamefont {I.~I.}\ \bibnamefont {Mazin}}, \bibinfo {author} {\bibfnamefont {L.}~\bibnamefont {Garcia-Gassull}}, \ and\ \bibinfo {author} {\bibfnamefont {R.}~\bibnamefont {Valent{\'\i}}},\ }\href@noop {} {\bibfield  {journal} {\bibinfo  {journal} {Physical Review B}\ }\textbf {\bibinfo {volume} {109}},\ \bibinfo {pages} {134424} (\bibinfo {year} {2024})}\BibitemShut {NoStop}%
\end{thebibliography}%

\end{document}